\documentclass[twocolumn,english,prl,superscriptaddress]{revtex4-1}
\usepackage[T1]{fontenc}
\usepackage[utf8]{inputenc}
\usepackage{array}
\usepackage{verbatim}
\usepackage{bm}
\usepackage{multirow}
\usepackage{amsmath}
\usepackage{amssymb}
\usepackage{graphicx}
\usepackage{rotating}
\usepackage{esint}

\makeatletter

\providecommand{\tabularnewline}{\\}
\usepackage{tikz}
\usetikzlibrary{calc}

\usepackage{braket}
\usepackage{CJK}

\usepackage{tikz}
\usepackage{pgfplots}
\usepackage{pgf}

\makeatother

\usepackage{babel}
\begin{document}
\begin{CJK*}{UTF8}{gbsn}
\title{Berry phase of the composite Fermi-liquid}
\author{Guangyue Ji (棘广跃)}
\affiliation{International Center for Quantum Materials, Peking University, Beijing
100871, China}
\author{Junren Shi (施均仁)}
\email{junrenshi@pku.edu.cn}

\affiliation{International Center for Quantum Materials, Peking University, Beijing
100871, China}
\affiliation{Collaborative Innovation Center of Quantum Matter, Beijing 100871,
China}
\begin{abstract}
We derive the definition of the Berry phase for the adiabatic transport
of a composite fermion (CF) in a half-filled composite Fermi-liquid
(CFL). It is found to be different from that adopted in previous investigations
by Geraedts et al. For the standard CFL wave function, we analytically
show that the Berry curvature is uniformly distributed in the momentum
space. For the Jain-Kamilla wave function, we numerically show that
its Berry curvature has a continuous distribution inside the Fermi
sea and vanishes outside. We conclude that the CF with respect to
both the microscopic wave-functions is not a massless Dirac particle.
\end{abstract}
\maketitle
\end{CJK*}

\section{Introduction\label{sec:Introduction}}

The ubiquitous presence of the Berry phase is notable in recent theoretical
investigations of condensed matter physics. For non-interacting systems,
it becomes a unifying concept for characterizing the orbital effects
of the spin or other internal degrees of freedom~\citep{xiao2010},
and plays central roles in systems such as topological insulators~\citep{hasan2010},
Dirac/Weyl semimetals~\citep{jia2016} and valleytronic materials~\citep{cao2012}.
Recently, it becomes clear that the Berry phase also plays a role
in the theory of composite Fermions (CFs)~\citep{jain2007}. CFs
can be regarded as weakly interacting particles residing in a hidden
Hilbert space~\citep{jain2009}. A wave function of non-interacting
CFs in the hidden Hilbert space can be mapped into a wave function
appropriate for describing the physical state of a strongly correlated,
fractionally filled Landau level. The  theory of CFs  achieves tremendous
successes in understanding the fractional quantum Hall effect and
related phenomena.~\citep{jain2007}.

However, although the wave functions prescribed by the CF theory are
shown to be very accurate and widely accepted~\citep{balram_nature_2016},
the effective theory of CFs interpreting the wave functions is still
open to debate. The conventional interpretation, as explicated in
Halperin-Lee-Read (HLR) theory of the composite Fermi-liquid (CFL)~\citep{kalmeyer1992,halperin1993,simon1998},
treats the CF as an ordinary Newtonian particle. In its pristine form,
it suffers from an apparent difficulty: it can not correctly predict
the CF Hall conductance of a half-filled Landau level~\citep{kivelson1997}.%
{} The difficulty motivates Son to propose that the CF should be a massless
Dirac particle~\citep{son2015}. An alternative interpretation, i.e.,
the CF is neither a Newtonian particle nor a Dirac particle, but a
particle subject to a uniformly distributed Berry curvature in the
momentum space and the Sundaram-Niu dynamics~\citep{sundaram1999},
is also put forward~\citep{haldane2016,shi2017,shi2018}. It is also
shown that the picture is equivalent to the dipole picture of CFs~\citep{read1994}.
The three pictures imply three different distributions of the Berry
curvature, i.e., zero, singularly distributed and uniformly distributed,
respectively. The clarification of the issue then hinges on the determination
of the Berry curvature for CFs.

A ``first principles'' approach for determining the Berry curvature
of CFs should be based on the microscopic CF wave-functions prescribed
by the CF theory. To this end, several attempts have been made. In
Ref.~\citep{shi2018}, the dynamics of the CF Wigner crystal is derived.
It shows that the CF is subject to a uniformly distributed Berry curvature
in the momentum space. For the half-filled CFL phase, the Berry curvature
distribution is  found to be uniform by determining the dynamics of
a test (distinguishable) CF added to the CF Fermi sea~\citep{shi2017}.
A heuristic argument based on the dipole picture of CFs also suggests
the same~\citep{haldane2016,shi2017}. These works may draw criticism
for neglecting the particle exchange symmetry in their treatments.
It is in this context that the recent works by Geraedts et al. stand
out~\citep{geraedts2018,wang_lattice_2019,wang_dirac_2019}. Their
calculations are based on a microscopic CFL wave function in its full
antisymmetric form. %
{} However, a close scrutiny to the works reveals a number of difficulties.
Firstly, the definition of the Berry phase is a prescribed one and
is not fully justified. Secondly, the evaluation of the Berry phase
based on the definition seems to be not numerically robust, sensitive
to the choices of paths and prone to statistical errors. Moreover,
there exist extraneous $\pm\pi/2$ phases preventing direct interpretations
of numerical results. Finally, the microscopic CFL wave function adopted
for the calculation is of the Jain-Kamilla (JK) type~\citep{jain1997a},
which is numerically efficient in implementing the projection to the
 Landau level (LLL). However, it is unclear whether or not it yields
the same result as that from the standard CFL wave function prescribed
by the theory of CFs~\citep{jain2007}.

In this paper, we solve these issues and determine the distribution
of the Berry curvature for CFs. First, we derive the definition of
the Berry phase directly from the original definition of the Berry
phase. It is found to be different from the prescribed one adopted
by Geraedts et al.~\citep{geraedts2018,wang_lattice_2019}. Then,
we analytically show that the Berry curvature distribution is uniform
in the momentum space for the standard CFL wave function. On the other
hand, to compare with the results in Refs.~\citep{geraedts2018,wang_lattice_2019},
we also evaluate the Berry curvature  of the JK wave function. With
our definition, the numerical evaluation of the Berry phase becomes
robust and free of the extraneous phases observed in Refs.~\citep{geraedts2018,wang_lattice_2019}.
 It enables us to numerically determine the distribution of the Berry
curvature in the whole momentum space. We find that the Berry curvature
has a continuous distribution inside the Fermi sea and vanishes outside,
and is different from the uniform distribution of the standard CFL
wave function. We analytically show that the difference originates
from the different quasi-periodicities of the two wave functions in
the reciprocal space.

The reminder is organized as follows. In Sec.~\ref{sec:definition_Berry_phase},
we derive the definition of the Berry phase of the CFL. In Sec.~\ref{sec:CF},
we determine the Berry phase and Berry curvature  of the standard
CFL wave function analytically. In Sec.~\ref{sec:JK}, we evaluate
the Berry curvature  of the JK wave function numerically. In Sec.~\ref{sec:Uniform-background},
we analyze the quasi-periodicities  of the two wave functions in the
reciprocal space, and determine the uniform background of the Berry
curvature. In Sec.~\ref{sec:Summary}, we summarize and discuss our
results.

\section{Definition of the Berry phase of CFL\label{sec:definition_Berry_phase}}

In this section, we derive the definition of the Berry phase for CFL
systems. We first introduce the generic definition of the Berry phase.
Next, we derive the definition of the Berry phase for CFL systems
from the generic definition. Then, we discuss different representations
of the Berry phases. Finally, we discuss and interpret Geraedts et
al.'s definition and results.

\subsection{Berry phase}

A quantum system acquires a geometric phase, i.e., the Berry phase,
when it is adiabatically transported along a path $C$ by varying
parameters $\bm{\alpha}$ in its Hamiltonian $\hat{H}\left(\bm{\alpha}\right)$~\citep{berry1984}.
The Berry phase is determined by a line integral in the parameter
space
\begin{align}
\gamma\left(C\right) & =\int_{C}\mathrm{i}\Braket{\Psi_{\bm{\alpha}}|\nabla_{\bm{\alpha}}\Psi_{\bm{\alpha}}}\cdot\mathrm{d}\bm{\alpha},\label{eq:gammaC}
\end{align}
where $\ket{\Psi_{\bm{\alpha}}}$ is the eigenstate of $\hat{H}(\bm{\alpha})$,
and the integrand is called the Berry connection. The phase is independent
of how the path is traversed as long as it is slow enough for the
adiabaticity to hold. For a closed path, the phase is gauge invariant,
i.e., independent of the choice of the phase factor of the wave function.

The Berry phase formula Eq.~(\ref{eq:gammaC}) can be recast into
an alternative form as a time integral:
\begin{equation}
\gamma\left(C\right)=\mathrm{i}\int_{t_{0}}^{t_{1}}\Braket{\Psi_{\bm{\alpha}(t)}|\frac{\mathrm{d}\Psi_{\bm{\alpha}(t)}}{\mathrm{d}t}}\mathrm{d}t,\label{eq:gammaCt}
\end{equation}
where the wave function evolves with time via the time-dependence
of its parameters, and $\bm{\alpha}(t)$ is an arbitrary time-dependent
function that traverses the path $C$ with $t_{0}$ ($t_{1}$) being
the beginning (ending) time of the evolution. The integrand is actually
a part of the Schr\"{o}dinger Lagrangian~\citep{sundaram1999,kramer1981}:
\begin{equation}
L=\Braket{\Psi(t)|\mathrm{i}\hbar\frac{\mathrm{d}}{\mathrm{d}t}-\hat{H}|\Psi(t)},
\end{equation}
which governs the time evolution of a quantum system. This is why
one sees the ubiquitous presence of the Berry phase in various contexts
such as effective dynamics~\citep{sundaram1999,kramer1981} and path-integral
formalisms.

\subsection{Definition of the Berry phase of CFL\label{subsec:Definition-of-the}}

From the generic definition of the Berry phase, we can infer a definition
of the Berry phase appropriate for CFL systems. For the purpose, it
is more convenient to first consider a much simpler system, i.e.,
a set of non-interacting electrons residing in a Bloch band. The definition
of the Berry phase for such a system is well known in the single-particle
form~\citep{sundaram1999}. Here, we will treat the system as a many-particle
system and find a many-body generalization of the Berry phase definition.
The generalization turns out to be general enough for applying to
CFL systems.

The many-body wave function of a set of non-interacting Bloch electrons
is a Slater determinant of Bloch states:
\begin{equation}
\Psi_{\bm{k}}\left(\bm{z}\right)=\mathrm{det}\left[\psi_{\bm{k}_{j}}\left(\bm{z}_{i}\right)\right],\label{eq:PsiBloch}
\end{equation}
where $\bm{k}\equiv\{\bm{k}_{1},\bm{k}_{2},\dots\}$ denotes the list
of the quasi-wave-vectors of the Bloch states occupied by electrons,
and $\psi_{\bm{k}_{j}}(\bm{z}_{i})=\exp(\mathrm{i}\bm{k}_{j}\cdot\bm{z}_{i})u_{\bm{k}_{j}}(\bm{z}_{i})$
is the Bloch wave function with $u_{\bm{k}_{j}}(\bm{z}_{i})$ being
its periodic part.

The wave function has a number of general properties which are actually
shared by the much more complicated CFL wave functions: (a) it is
parameterized by a set of wave-vectors $\bm{k}$; (b) it is an eigenstate
of the (magnetic) center-of-mass translation operator $\hat{T}\left(\bm{a}\right)$
such that $\hat{T}(\bm{a})\Psi_{\bm{k}}(\bm{z})=\exp(\mathrm{i}\sum_{i}\bm{k}_{i}\cdot\bm{a})\Psi_{\bm{k}}(\bm{z})$,
where $\bm{a}$ is one of the vectors of the Bravais lattice with
respect to the periodicity of the system. As a result, two states
with different total wave-vectors are orthogonal to each other. With
proper normalizations of wave functions, we have:
\begin{equation}
\Braket{\Psi_{\bm{k}}|\Psi_{\bm{k}^{\prime}}}=\delta\left(\sum_{i}\bm{k}_{i}-\sum_{i}\bm{k}_{i}^{\prime}\right)f\left(\bm{k},\bm{k}^{\prime}\right),\label{eq:normalization}
\end{equation}
where $f(\bm{k},\bm{k}^{\prime})$ is a function with the property
$f(\bm{k},\bm{k})=1$, and the wave-vectors in the Dirac Delta function
are regarded equal if they are only different by a reciprocal lattice
vector; (c) the wave function has the Fermionic exchange symmetry
and can be obtained from an unsymmetrized wave function $\varphi_{\bm{k}}$
by applying the anti-symmetrization operator $\hat{\mathcal{P}}$:
\begin{align}
\hat{\mathcal{P}} & =\frac{1}{N!}\sum_{P}\left(-1\right)^{P}\hat{P},\\
\Psi_{\bm{k}}\left(\bm{z}\right) & =\hat{\mathcal{P}}\varphi_{\bm{k}}(\bm{z})\equiv\frac{1}{N!}\sum_{P}\left(-1\right)^{P}\varphi_{\bm{k}}(\hat{P}\bm{z}),
\end{align}
where $\hat{P}\bm{z}$ denotes a permutation of electron coordinates,
$N$ is the total number of electrons, and $\varphi_{\bm{k}}(\bm{z})=\sqrt{N!}\prod_{i}\psi_{\bm{k}_{i}}(\bm{z}_{i})$
for the Bloch system. In the unsymmetrized form, an electron is associated
with a particular wave-vector. The association is lost in the antisymmetrized
form.

With the wave function in hand, one may be tempted to directly apply
Eq.~(\ref{eq:gammaC}) to determine the Berry phase. However, a difficulty
immediately arises. To see that, we treat $\bm{k}_{1}$ as the parameters
$\bm{\alpha}$, substitute Eq.~(\ref{eq:PsiBloch}) into Eq.~(\ref{eq:gammaC}),
apply the identity $\hat{\mathcal{P}}^{2}=\hat{\mathcal{P}}$, and
obtain $\bm{A}_{\bm{k}_{1}}\equiv\mathrm{i}\braket{\Psi_{\bm{k}}|\partial_{\bm{k}_{1}}\Psi_{\bm{k}}}=-\braket{\Psi_{\bm{k}}|\bm{r}_{1}|\varphi_{\bm{k}}}+\mathrm{i}\braket{\Psi_{\bm{k}}|e^{\mathrm{i}\bm{k}_{1}\cdot\bm{r}_{1}}|\partial_{\bm{k}_{1}}u_{\bm{k}_{1}}(\bm{r}_{1})\prod_{i\geq2}\psi_{\bm{k}_{i}}(\bm{r}_{i})}$.
Unfortunately, the resulting Berry connection $\bm{A}_{\bm{k}_{1}}$
is not a legitimate one because $\braket{\Psi_{\bm{k}}|\bm{r}_{1}|\varphi_{\bm{k}}}$
does not have a deterministic value since the Bloch states have definite
momenta and therefore infinite position uncertainty. This is the difficulty
we have to address before the generic definition can be applied to
wave functions like Eq.~(\ref{eq:PsiBloch}).

The most straightforward approach to address the issue is to introduce
a unitary transformation to the Hamiltonian: $\hat{H}(\bm{k}_{1})=e^{-\mathrm{i}\bm{k}_{1}\cdot\bm{r}_{1}}\hat{H}e^{\mathrm{i}\bm{k}_{1}\cdot\bm{r}_{1}}$.
The resulting Hamiltonian acquires dependence on the parameters $\bm{k}_{1}$,
and the corresponding eigenstate wave function becomes $e^{-\mathrm{i}\bm{k}_{1}\cdot\bm{r}_{1}}\varphi_{\bm{k}}(\bm{r})=u_{\bm{k}_{1}}(\bm{r}_{1})\prod_{i\geq2}\psi_{\bm{k}_{i}}(\bm{r}_{i})$.
One can then apply Eq.~(\ref{eq:gammaC}) to obtain the well-known
result $\bm{A}_{\bm{k}_{1}}=\mathrm{i}\braket{u_{\bm{k}_{1}}|\partial_{\bm{k}_{1}}u_{\bm{k}_{1}}}$.
Such an approach is adopted and generalized in Ref.~\citep{shi2017}
to show that the standard CFL wave function yields a uniform Berry
curvature $\Omega(\bm{k}_{1})=1/qB$, where $q$ is the unit charge
of carriers and $B$ is the perpendicular component of the external
magnetic field $\bm{B}$. However, the approach is not compatible
with the exchange symmetry because $\hat{H}(\bm{k}_{1})$ obviously
breaks the symmetry of exchanging the first particle (the particle
being transported) with others. Adopting such an approach means that
we have to ignore the exchange symmetry. This is what we want to avoid
here.

We therefore adopt and generalize the approach presented in Ref.~\citep{sundaram1999}.
The basic idea is that, since the difficulty is due to the fact that
a Bloch state does not have a deterministic position expectation value,
we replace it with a wave packet state which has a central wave-vector
$\bm{k}_{c}$ and give rises to a deterministic position expectation
value $\bm{z}_{c}$:
\begin{equation}
\text{\ensuremath{\Ket{\tilde{\Psi}_{\bm{k}_{c},\bm{z}_{c}}}=\int\mathrm{d}\bm{k}_{1}a\left(\bm{k}_{1},t\right)\Ket{\Psi_{\bm{k}}},}}\label{eq:psitilde}
\end{equation}
where we assume that $|a(\bm{k}_{1},t)|^{2}$ is narrowly distributed
around $\bm{k}_{c}$ and satisfies
\begin{align}
 & \int\mathrm{d}\bm{k}_{1}\left|a\left(\bm{k}_{1},t\right)\right|^{2}=\Braket{\tilde{\Psi}_{\bm{k}_{c},\bm{z}_{c}}|\tilde{\Psi}_{\bm{k}_{c},\bm{z}_{c}}}=1,\\
 & \int\mathrm{d}\bm{k}_{1}\left|a\left(\bm{k}_{1},t\right)\right|^{2}\bm{k}_{1}=\bm{k}_{c}.
\end{align}
We choose the time-dependence of $a\left(\bm{k}_{1},t\right)$ to
make $\bm{k}_{c}$ traverses a path $C$ while keeping $\bm{z}_{c}$
fixed. By applying Eq.~(\ref{eq:gammaCt}), we can then determine
the Berry phase acquired by the wave-packet state. In the end, the
width of the distribution $|a(\bm{k}_{1},t)|^{2}$ will be set to
zero so that the wave-packet state approaches to the Bloch state.
We will show that it yields a well-defined limit.

We still need to define $\bm{z}_{c}$. It is easy to see that $\braket{\tilde{\Psi}_{\bm{k}_{c},\bm{z}_{c}}|\bm{z}_{1}|\tilde{\Psi}_{\bm{k}_{c},\bm{z}_{c}}}$
does not yield a deterministic expectation value. This is because
$\bm{z}_{1}$ loses its association with $\bm{k}_{1}$ in the antisymmetrized
wave function, and $\braket{\tilde{\Psi}_{\bm{k}_{c},\bm{z}_{c}}|\bm{z}_{1}|\tilde{\Psi}_{\bm{k}_{c},\bm{z}_{c}}}=\braket{\tilde{\Psi}_{\bm{k}_{c},\bm{z}_{c}}|\bm{z}_{i}|\tilde{\Psi}_{\bm{k}_{c},\bm{z}_{c}}}$
is nothing but the center-of-mass position. Since electrons, all but
one, have definite wave-vectors in $\ket{\tilde{\Psi}_{\bm{k}_{c},\bm{z}_{c}}}$,
the center-of-mass position has infinite uncertainty. To obtain a
deterministic position, we define $\bm{z}_{c}$ as the position of
the electron associated with the wave-vector $\bm{k}_{1}$ by: 
\begin{equation}
\bm{z}_{c}=\mathrm{Re}\Braket{\tilde{\Psi}_{\bm{k}_{c},\bm{z}_{c}}|\bm{z}_{1}|\tilde{\varphi}_{\bm{k}_{c},\bm{z}_{c}}},\label{eq:zc}
\end{equation}
where $\ket{\tilde{\varphi}_{\bm{k}_{c},\bm{z}_{c}}}$ is the unsymmetrized
form of $\ket{\tilde{\Psi}_{\bm{k}_{c},\bm{z}_{c}}}$, i.e., $\ket{\tilde{\Psi}_{\bm{k}_{c},\bm{z}_{c}}}\equiv\hat{\mathcal{P}}\ket{\tilde{\varphi}_{\bm{k}_{c},\bm{z}_{c}}}$. 

We can show that $\bm{z}_{c}$ does have a deterministic value. To
see this, we substitute Eq.~(\ref{eq:psitilde}) into Eq.~(\ref{eq:zc}),
and have
\begin{align}
\bm{z}_{c}= & \mathrm{Re}\int\mathrm{d}\bm{k}_{1}\int\mathrm{d}\bm{k}_{1}^{\prime}a^{\ast}\left(\bm{k}_{1}^{\prime},t\right)a\left(\bm{k}_{1},t\right)\nonumber \\
 & \times\left[-\mathrm{i}\partial_{\bm{k}_{1}}\Braket{\Psi_{\bm{k}^{\prime}}|\varphi_{\bm{k}}}+\mathrm{i}\Braket{\Psi_{\bm{k}^{\prime}}|e^{\mathrm{i}\bm{k}_{1}\cdot\bm{z}_{1}}|\partial_{\bm{k}_{1}}u_{\bm{k}}}\right]\nonumber \\
= & \mathrm{Re}\int\mathrm{d}\bm{k}_{1}[a^{\ast}\left(\bm{k}_{1},t\right)\mathrm{i}\partial_{\bm{k}_{1}}a\left(\bm{k}_{1},t\right)\nonumber \\
 & +\mathrm{i}\left|a\left(\bm{k}_{1},t\right)\right|^{2}\Braket{\Psi_{\bm{k}}|e^{\mathrm{i}\bm{k}_{1}\cdot\bm{z}_{1}}|\partial_{\bm{k}_{1}}u_{\bm{k}}}],
\end{align}
where we define $u_{\bm{k}}(\bm{z})\equiv e^{-\mathrm{i}\bm{k}_{1}\cdot\bm{z}_{1}}\varphi_{\bm{k}}(\bm{z})$.
To obtain the last expression, we make use of Eq.~(\ref{eq:normalization})
which reduces to $\braket{\Psi_{\bm{k}^{\prime}}|\varphi_{\bm{k}}}=\braket{\Psi_{\bm{k}^{\prime}}|\Psi_{\bm{k}}}=\delta\left(\bm{k}_{1}-\bm{k}_{1}^{\prime}\right)$
for the current case. Moreover, one can show that $e^{\mathrm{i}\bm{k}_{1}\cdot\bm{z}_{1}}\ket{\partial_{\bm{k}_{1}}u_{\bm{k}}}$
has the same center-of-mass periodicity as $\ket{\Psi_{\bm{k}}}$,
thus $\Braket{\Psi_{\bm{k}^{\prime}}|e^{\mathrm{i}\bm{k}_{1}\cdot\bm{z}_{1}}|\partial_{\bm{k}}u_{\bm{k}}}\propto\delta(\bm{k}_{1}-\bm{k}_{1}^{\prime})$.

Writing the amplitude $a\left(\bm{k}_{1},t\right)$ as the form $a\left(\bm{k}_{1},t\right)=\left|a\left(\bm{k}_{1},t\right)\right|e^{-\mathrm{i}\gamma(\bm{k}_{1},t)}$,
and setting the width of the distribution $\left|a(\bm{k}_{1},t)\right|^{2}$
to zero, we obtain
\begin{equation}
\bm{z}_{c}=\frac{\partial\gamma\left(\tilde{\bm{k}},t\right)}{\partial\bm{k}_{c}}-\mathrm{Im}\Braket{\Psi_{\tilde{\bm{k}}}|e^{\mathrm{i}\bm{k}_{c}\cdot\bm{z}_{1}}|\partial_{\bm{k}_{c}}u_{\tilde{\bm{k}}}}\label{eq:zc-1}
\end{equation}
with $\tilde{\bm{k}}\equiv\{\bm{k}_{c.},\bm{k}_{2},\dots\}$. Equation
(\ref{eq:zc-1}) is the many-body generalization of Eq.~(2.8) of
Ref.~\citep{sundaram1999}.

We then apply Eq.~(\ref{eq:gammaCt}) to determine the Berry phase
of transporting $\bm{k}_{c}$. Following a procedure similar to that
for determining $\bm{z}_{c}$, we obtain
\begin{align}
\mathrm{i}\Braket{\tilde{\Psi}_{\bm{k}_{c}(t),\bm{z}_{c}}|\frac{\mathrm{d}\tilde{\Psi}_{\bm{k}_{c}(t),\bm{z}_{c}}}{\mathrm{d}t}} & =\frac{\partial\gamma\left(\tilde{\bm{k}},t\right)}{\partial t}.\label{eq:dPsidt}
\end{align}
Using the relation
\begin{equation}
\frac{\partial\gamma\left(\tilde{\bm{k}},t\right)}{\partial t}=\frac{\mathrm{d}\gamma\left(\tilde{\bm{k}},t\right)}{\mathrm{d}t}-\dot{\bm{k}}_{c}\cdot\frac{\partial\gamma\left(\tilde{\bm{k}},t\right)}{\partial\bm{k}_{c}}
\end{equation}
and Eq.~(\ref{eq:zc-1}), we determine the phase acquired by transporting
$\bm{k}_{c}$ along $C$:
\begin{multline}
\gamma\left(C\right)=\int_{t_{0}}^{t_{1}}\mathrm{d}t\left[\frac{\mathrm{d}\gamma\left(\tilde{\bm{k}},t\right)}{\mathrm{d}t}-\dot{\bm{k}}_{c}\cdot\bm{z}_{c}\right]\\
-\int_{C}\mathrm{Im}\Braket{\Psi_{\tilde{\bm{k}}}|e^{\mathrm{i}\bm{k}_{c}\cdot\bm{z}_{1}}|\partial_{\bm{k}_{c}}u_{\tilde{\bm{k}}}}\cdot\mathrm{d}\bm{k}_{c}.\label{eq:gammac1}
\end{multline}
The first term is integrable (provided $\bm{z}_{c}$ is fixed) and
vanishes when $C$ is a closed path. We thus exclude it from the definition
of the Berry phase, and define the Berry connection as:
\begin{equation}
\bm{A}_{\bm{k}_{1}}=-\mathrm{Im}\Braket{\Psi_{\bm{k}}|e^{\mathrm{i}\bm{k}_{1}\cdot\bm{z}_{1}}|\partial_{\bm{k}_{1}}u_{\bm{k}}},\label{eq:Ak1}
\end{equation}
where we relabel $\bm{k}_{c}$ as $\bm{k}_{1}$. The Berry phase is
just a line integral of the Berry connection.

For numerical calculations, it is more convenient to calculate the
Berry phase induced by a small but discrete change of the wave-vector.
By using the trapezoidal rule to estimate the line integral and approximating
$\partial_{\bm{k}_{1}}u_{\bm{k}}$ as a first-order divided difference,
we determine the Berry phase for $\bm{k}_{1}\rightarrow\bm{k}_{1}^{\prime}$:
\begin{equation}
\phi_{\bm{k}^{\prime}\bm{k}}=\frac{1}{2}\left[\mathrm{arg}\Braket{\varphi_{\bm{k}^{\prime}}|e^{\mathrm{i}\bm{q}\cdot\bm{z}_{1}}|\Psi_{\bm{k}}}-(\bm{k}\rightleftharpoons\bm{k}^{\prime})\right],\label{eq:BerryPhase}
\end{equation}
with $\bm{q}\equiv\bm{k}_{1}^{\prime}-\bm{k}_{1}$.

Equation (\ref{eq:Ak1}) and (\ref{eq:BerryPhase}) are the definitions
of the Berry connection and Berry phase for a many-body system, respectively.
The definitions are directly descended from the original definition
Eq.~(\ref{eq:gammaC}). In our derivation, we only make use of the
aforementioned three properties of the wave function. As we will show
later, CFL wave functions considered in this paper do have these properties.
Therefore, the definitions are also applicable for CFL systems.

\subsection{Representations of the Berry phase}

Motivated by the dipole picture of CFs~\citep{read1994}, we also
define another Berry phase (connection). According to the dipole picture,
a CF is a bound state of an electron and quantum vortices, and the
position of the quantum vortices $\bm{z}_{c}^{\mathrm{v}}$ is displaced
from that of the electron by~\citep{shi2017,shi2018}
\begin{equation}
\bm{z}_{c}^{\mathrm{v}}=\bm{z}_{c}^{\mathrm{e}}+\bm{n}\times\bm{k}_{c},\label{eq:zcv}
\end{equation}
where $\bm{z}_{c}^{\mathrm{e}}\equiv\bm{z}_{c}$ is the position of
the electron, and $\bm{n}$ denotes the unit normal vector of the
system plane. Obviously, the phase determined by Eq.~(\ref{eq:gammac1})
depends on which position is fixed when $\bm{k}_{c}$ is transported.
The Berry connection Eq.~(\ref{eq:Ak1}) and the Berry phase Eq.~(\ref{eq:BerryPhase})
are defined by assuming that $\bm{z}_{c}^{\mathrm{e}}$ is fixed.
Hereafter, we will label $\bm{A}_{\bm{k}_{1}}$ ($\phi_{\bm{k}^{\prime}\bm{k}}$)
as $\bm{A}_{\bm{k}_{1}}^{\mathrm{e}}$ ($\phi_{\bm{k}^{\prime}\bm{k}}^{\mathrm{e}}$)
to explicitly indicate the assumption. On the other hand, if we assume
that $\bm{z}_{c}^{\mathrm{v}}$ is fixed, we should replace $\bm{z}_{c}$
in Eq.~(\ref{eq:gammac1}) with $\bm{z}_{c}^{\mathrm{v}}-\bm{n}\times\bm{k}_{c}$,
and obtain another Berry connection $\bm{A}_{\bm{k}_{1}}^{\mathrm{v}}$:
\begin{equation}
\bm{A}_{\bm{k}_{1}}^{\mathrm{v}}=\bm{A}_{\bm{k}_{1}}^{\mathrm{e}}-\bm{k}_{1}\times\bm{n}.
\end{equation}
The corresponding Berry phase for $\bm{k}_{1}\rightarrow\bm{k}_{1}^{\prime}$
is
\begin{equation}
\phi_{\bm{k}^{\prime}\bm{k}}^{\mathrm{v}}=\phi_{\bm{k}^{\prime}\bm{k}}^{\mathrm{e}}+(\bm{k}_{1}\times\bm{k}_{1}^{\prime})\cdot\bm{n}.\label{eq:phiv}
\end{equation}

Then, which one is the Berry connection (phase) of the CF? The answer
depends on how we define the  position of a CF. By definition, the
$\bm{k}$-space Berry connection is the connection of transporting
$\bm{k}_{1}$ while keeping the position fixed. If one chooses to
define the CF position as the position of the quantum vortices (electron),
then $\bm{A}_{\bm{k}_{1}}^{\mathrm{v}}$ ($\bm{A}_{\bm{k}_{1}}^{\mathrm{e}}$)
should be regarded as the CF Berry connection. One can even adopt
other definitions of the CF position, and obtain other definitions
of the Berry connections. Mathematically, all these definitions are
 equivalent. They are just different representations of the same physics.

Nevertheless, for a reason to be discussed in the next subsection,
we will call the v-representation as the CF representation, and interpret
$\phi_{\bm{k}^{\prime}\bm{k}}^{\mathrm{v}}$ and $\bm{A}_{\bm{k}_{1}}^{\mathrm{v}}$
as the Berry phase and Berry connection of CFs.

\subsection{Interpretation of Geraedts et al.'s result\label{subsec:Interpretation-of-Geraedts}}

Finally, we would like to comment on the definition of the Berry phase
introduced by Geraedts et al. in Refs.~\citep{geraedts2018,wang_lattice_2019}.
It is obviously not a definition descended from the original definition
of the Berry phase. It should be more appropriately called as a scattering
phase. Indeed, the transition amplitude for $\bm{k}\rightarrow\bm{k}^{\prime}$
induced by a single-body scalar potential $V(\bm{z}_{i})=e^{\mathrm{i}\bm{q}\cdot\bm{z}_{i}}$
is proportional to
\begin{equation}
U_{\bm{k}^{\prime}\bm{k}}=\Braket{\Psi_{\bm{k}^{\prime}}|\sum_{i}e^{\mathrm{i}\bm{q}\cdot\bm{z}_{i}}|\Psi_{\bm{k}}}
\end{equation}
with $\bm{k}_{1}^{\prime}=\bm{k}_{1}+\bm{q}$. It is exactly the matrix
element calculated by Geraedts et al. Their results thus provide a
``first-principles'' determination of the scattering amplitude. 

We can actually interpret Geraedts et al.'s results in light of the
picture of independent CFs. They observe that the matrix element has
an $\mathrm{i}$ ($-\mathrm{i}$) factor for a discrete wave-vector
change in the clockwise (anti-clockwise) sense~\footnote{ The correspondence is determined from our own numerical evaluation
of the matrix element. }. It means that the matrix element has a form like
\begin{equation}
U_{\bm{k}^{\prime}\bm{k}}\propto\mathrm{i}\left(\bm{k}_{1}^{\prime}\times\bm{k}_{1}\right)e^{\mathrm{i}\Phi_{\bm{k}^{\prime}\bm{k}}},
\end{equation}
and $\Phi_{\bm{k}^{\prime}\bm{k}}$ is interpreted as the Berry phase.
For a CF system, the potential will induce a density modulation, which
in turn induces a modulation of the effective magnetic field. The
scattering amplitude induced by the modulated effective-magnetic-field
does have a factor $\propto\mathrm{i}\left(\bm{k}_{1}^{\prime}\times\bm{k}_{1}\right)$,
as shown in Eq.~(23) of Ref.~\citep{wang_particle-hole_2017}. Different
from what is assumed in Ref.~\citep{wang_particle-hole_2017}, the
scattering amplitude carries an extra phase $\Phi_{\bm{k}^{\prime}\bm{k}}$,
indicating the presence of a Berry phase when the transition $\bm{k}\rightarrow\bm{k}^{\prime}$
occurs. 

It is then interesting to ask how the Berry phase inferred from the
scattering amplitude is related to the Berry phases we have defined.
Our numerical calculation (See Sec.~\ref{subsec:Berry-curvature-distribution})
suggests that
\begin{equation}
\Phi_{\bm{k}^{\prime}\bm{k}}\approx\phi_{\bm{k}^{\prime}\bm{k}}^{\mathrm{v}},\label{eq:Phiphi}
\end{equation}
i.e., the Berry phase inferred from the scattering amplitude is actually
the Berry phase defined in the v-representation. For the dipole picture,
it means that as far as the impurity scattering is concerned, the
position of a CF should be defined as the position of its quantum
vortices. 

On the other hand, one expects that CFs are scattered not only by
the fluctuation of the effective magnetic field, but also by the scalar
potential itself.%
{} As a result, the scattering amplitude should in general have the
form~\citep{wang_particle-hole_2017}:
\begin{equation}
U_{\bm{k}^{\prime}\bm{k}}=\left[V_{1}\left(\bm{k}^{\prime},\bm{k}\right)+\mathrm{i}\left(\bm{k}_{1}^{\prime}\times\bm{k}_{1}\right)V_{2}\left(\bm{k}^{\prime},\bm{k}\right)\right]e^{\mathrm{i}\Phi_{\bm{k}^{\prime}\bm{k}}}.\label{eq:U}
\end{equation}
The presence of the first term will interfere the determination of
$\Phi_{\bm{k}^{\prime}\bm{k}}$. Indeed, we observe that the Berry
phase inferred from the scattering amplitude by assuming a vanishing
$V_{1}\left(\bm{k}^{\prime},\bm{k}\right)$ always deviates from $\pi$,
and the deviation becomes more pronounced when $N$ is scaled up (see
Table~\ref{table2}). %
The approach becomes unreliable when the two terms in the prefactor
of Eq.~(\ref{eq:U}) are comparable in magnitude. In this case, the
phase carried by the prefactor cannot be easily distinguished from
the Berry phase to be determined. One encounters such a situation
when studying the effect of the Landau level mixing~\citep{pu_berry_2018}.
In the study, Pu et al. adopt the wave function
\begin{equation}
\Psi_{\bm{k}}^{\mathrm{mix}}=\left(1-\beta\right)\Psi_{\bm{k}}+\beta\Psi_{\bm{k}}^{\mathrm{unproj.}},
\end{equation}
where $\Psi_{\bm{k}}$ is a CFL wave function projected to the LLL
(either Eq.~(\ref{eq:CFL}) or Eq.~(\ref{eq:JK})), $\Psi_{\bm{k}}^{\mathrm{unproj.}}$
is the unprojected form of $\Psi_{\bm{k}}$ (i.e., Eq.~(\ref{eq:CFL})
without applying $\hat{P}_{\mathrm{LLL}}$), and $\beta$ controls
the strength of the Landau level mixing. One expects that $V_{1}$
dominates in the scattering amplitude of the unprojected wave function
$\Psi_{\bm{k}}^{\mathrm{unproj.}}$, and $V_{1}>0$ ($V_{1}<0$) for
particles (holes). From Eq.~(\ref{eq:U}), the prefactor contributes
a phase $0$ ($\pi$) for each step of transporting a particle (hole)~\citep{pu_berry_2018}.
On the other hand, for the projected wave function $\Psi_{\bm{k}}$,
$V_{2}$ dominates, and the prefactor contributes a phase $\pm\pi/2$.
In between for a mixed wave function, the phase depends on the relative
strength of $V_{1}$ and $V_{2}$. Without a reliable way of subtracting
the phase from the scattering amplitude, the determined ``Berry phase''
may fluctuate widely. This is indeed observed in Ref.~\citep{pu_berry_2018}.
In contrast, our definition Eq.~(\ref{eq:BerryPhase}) does not suffer
from the difficulty. It is easy to see that the Berry phase of $\Psi_{\bm{k}}^{\mathrm{unproj.}}$
is always zero. For the mixed wave function, it is reasonable to expect
that the Berry phase accumulated by transporting a CF around the Fermi
circle is a value between zero and that yielded by $\Psi_{\bm{k}}$,
i.e., $\pi$.

\section{Berry phase of the standard CFL wave function\label{sec:CF}}

In this section, we evaluate the Berry phase of the standard CFL wave
function. First, we introduce the explicit form of the CFL wave function
on the torus geometry. Next, we show its center-of-mass translational
symmetry under the magnetic translation operator. Then, we analytically
determine the Berry curvature  of the standard CFL wave function.

To unify notations, we use the symbols $a_{i}\equiv a_{ix}+\mathrm{i}a_{iy}$,
$a_{i}^{\ast}\equiv a_{ix}-\mathrm{i}a_{iy}$ and $\bm{a}_{i}\equiv(a_{ix},a_{iy})$
to denote an electron-related variable in its complex form, complex
conjugate and vector form, respectively, with the subscript $i$ indexing
electrons. Symbols without a subscript (e.g. $a\equiv\{a_{i}\}$)
denote a list of the variables for all electrons, and symbols in the
upper case (e.g. $A\equiv\sum_{i}a_{i}$) denote sums of the variables
over all electrons. $a\cdot b\equiv\sum_{i}a_{i}b_{i}$ denotes the
inner product of two lists of variables. The unit of length is set
to the magnetic length $l_{B}\equiv\sqrt{\hbar/e|B|}$.

\subsection{CFL wave function\label{subsec:Wave-function}}

The standard CFL wave function for a system on a torus with a filling
fraction $\nu=1/m$ ($m$ is an even integer)  can be written as (omitting
the Gaussian factor $e^{-\sum_{i}\left|z_{i}\right|^{2}/4}$)~\citep{shao2015}
\begin{align}
\Psi_{\bm{k}}^{\mathrm{CF}}\left(z\right) & =\hat{P}_{\mathrm{LLL}}\mathrm{det}\left[e^{\mathrm{i}\left(k_{i}z_{j}^{\ast}+k_{i}^{\ast}z_{j}\right)/2}\right]J(z),\label{eq:CFL}\\
J(z) & =\tilde{\sigma}^{m}\left(Z\right)\prod_{i<j}\tilde{\sigma}^{m}(z_{i}-z_{j}),\label{eq:Jastrow}
\end{align}
 where $\hat{P}_{\mathrm{LLL}}$ denotes the projection to the LLL,
which is effectively to replace $z_{i}^{\ast}$ with an operator $2\partial_{z_{i}}$
acting on all $z_{i}$'s~\citep{jain2007},   and $J(z)$ is the
Bijl-Jastrow factor~\citep{jain2009} expressed in terms of the modified
sigma function $\tilde{\sigma}$ which has the quasi-periodicity~\citep{haldane2018}
\begin{equation}
\tilde{\sigma}(z_{i}+L)=\xi(L)e^{\frac{\pi L^{\ast}}{A}\left(z_{i}+\frac{1}{2}L\right)}\tilde{\sigma}(z_{i}),\label{eq:sigma}
\end{equation}
where $L$ is a period of the torus, and $\xi(L)=1$ if $L/2$ is
also a period and $-1$ otherwise, $A=2\pi N_{\phi}\equiv2\pi mN$
is the total area of the torus, and $N_{\phi}$ is the total number
of the magnetic fluxes passing through the torus. The wave function
is parameterized in a set of wave vectors $\bm{k}$ which are quantized
as usual, i.e., $\bm{k}_{i}\in\{n_{1}\bm{b}_{1}+n_{2}\bm{b}_{2},n_{1},n_{2}\in Z\}$
with $\bm{b}_{\alpha}=(2\pi/A)\bm{n}\times\bm{L}_{\alpha}$, $\alpha=1,2$,
where $\bm{L}_{1}$ and $\bm{L}_{2}$ denote the two edges defining
the torus. In the complex form, we have $k_{i}\in\{(n_{1}L_{1}+n_{2}L_{2})/N_{\phi},n_{1},n_{2}\in Z\}$.

After applying the projection and expanding the determinant, we write
the wave function explicitly as $\Psi_{\bm{k}}^{\mathrm{CF}}\left(z\right)=\hat{\mathcal{P}}\varphi_{\bm{k}}^{\mathrm{CF}}\left(z\right)$,
and
\begin{multline}
\varphi_{\bm{k}}^{\mathrm{CF}}\left(z\right)=N!e^{\mathrm{i}k^{\ast}\cdot z/2}\tilde{\sigma}^{m}\left(Z+\mathrm{i}K\right)\\
\times\prod_{i<j}\tilde{\sigma}^{m}(z_{i}+\mathrm{i}k_{i}-z_{j}-\mathrm{i}k_{j}).\label{eq:CFL2}
\end{multline}
 $\varphi_{\bm{k}}^{\mathrm{CF}}$ is the unsymmetrized form of $\Psi_{\bm{k}}^{\mathrm{CF}}$.

\subsection{Translational symmetry\label{subsec:Translational-symmetry}}

The wave function is an eigenstate of the magnetic center-of-mass
translation operator defined by~\citep{zak_magnetic_1964}
\begin{align}
\hat{T}\left(\bm{a}\right) & =\prod_{i}e^{\bm{a}\cdot\partial_{\bm{z}_{i}}+\frac{1}{2}\mathrm{i}\left(\bm{n}\times\bm{z}_{i}\right)\cdot\bm{a}}\nonumber \\
 & =e^{\frac{1}{4}\left(aZ^{\ast}-a^{\ast}Z\right)}\prod_{i}e^{\bm{a}\cdot\partial_{\bm{z}_{i}}}
\end{align}
and $\bm{a}\in\{n\bm{L}/N,n\in Z\}$, where $\bm{L}$ is a period
of the torus. To show that, we apply $\hat{T}(\bm{a})$ to the wave
function Eq.~(\ref{eq:CFL2}), and have 
\begin{multline}
\hat{T}\left(\bm{a}\right)\varphi_{\bm{k}}^{\mathrm{CF}}\left(z\right)e^{-\frac{\sum_{i}\left|z_{i}\right|^{2}}{4}}=e^{\frac{1}{2}\mathrm{i}K^{\ast}a-\frac{a^{\ast}}{2}\left(Z+\frac{nL}{2}\right)}\\
\times e^{\frac{1}{2}\mathrm{i}k^{\ast}\cdot z}\tilde{\sigma}^{m}\left(Z+\mathrm{i}K+nL\right)\\
\times e^{-\frac{\sum_{i}\left|z_{i}\right|^{2}}{4}}\prod_{i<j}\tilde{\sigma}^{m}(z_{i}+\mathrm{i}k_{i}-z_{j}-\mathrm{i}k_{j}),\label{eq:Tavarphi}
\end{multline}
where we assume $\bm{a}=n\bm{L}/N$, $n\in Z$. Using Eq.~(\ref{eq:sigma}),
we have 
\begin{equation}
\tilde{\sigma}^{m}\left(Z+iK+nL\right)=e^{\frac{1}{2}a^{\ast}\left(Z+\mathrm{i}K+\frac{nL}{2}\right)}\tilde{\sigma}^{m}\left(Z+iK\right).
\end{equation}
Substituting the relation into Eq.~(\ref{eq:Tavarphi}), and noting
that $\hat{T}(\bm{a})$ commutes with $\hat{\mathcal{P}}$, we obtain
\begin{equation}
\hat{T}\left(\bm{a}\right)\Psi_{\bm{k}}^{\mathrm{CF}}\left(z\right)e^{-\frac{\sum_{i}\left|z_{i}\right|^{2}}{4}}=e^{\mathrm{i}\bm{K}\cdot\bm{a}}\Psi_{\bm{k}}^{\mathrm{CF}}\left(z\right)e^{-\frac{\sum_{i}\left|z_{i}\right|^{2}}{4}}.
\end{equation}

We can also show that $\tilde{\varphi}_{\bm{k}}(\bm{z})\equiv e^{\mathrm{i}\bm{k}_{1}\cdot\bm{z}_{1}}\partial_{\bm{k}_{1}}u_{\bm{k}}(\bm{z})e^{-\frac{\sum_{i}\left|z_{i}\right|^{2}}{4}}$
with $u_{\bm{k}}(\bm{z})\equiv e^{-\mathrm{i}\bm{k}_{1}\cdot\bm{z}_{1}}\varphi_{\bm{k}}^{\mathrm{CF}}\left(z\right)$
is an eigenstate of $\hat{T}(\bm{a})$ of the same eigenvalue. It
is easy to obtain
\begin{equation}
\hat{T}\left(\bm{a}\right)u_{\bm{k}}\left(\bm{z}\right)e^{-\frac{\sum_{i}\left|z_{i}\right|^{2}}{4}}=e^{\mathrm{i}\sum_{i\geq2}\bm{k}_{i}\cdot\bm{a}}u_{\bm{k}}\left(\bm{z}\right)e^{-\frac{\sum_{i}\left|z_{i}\right|^{2}}{4}}.
\end{equation}
Using the relation, we obtain
\begin{align}
\hat{T}\left(\bm{a}\right)\tilde{\varphi}_{\bm{k}}(\bm{z}) & =e^{\mathrm{i}\bm{k}_{1}\cdot\left(\bm{z}_{1}+\bm{a}\right)}\partial_{\bm{k}_{1}}\hat{T}\left(\bm{a}\right)u_{\bm{k}}(\bm{z}))e^{-\frac{\sum_{i}\left|z_{i}\right|^{2}}{4}}\nonumber \\
 & =e^{\mathrm{i}\bm{K}\cdot\bm{a}}\tilde{\varphi}_{\bm{k}}(\bm{z}).
\end{align}
It is indeed an eigenstate of $\hat{T}(\bm{a})$ with an eigenvalue
$e^{\mathrm{i}\bm{K}\cdot\bm{a}}$.

\subsection{Berry phase and Berry curvature\label{subsec:CF}}

It is obvious from the above discussions that the standard CFL wave
function have all the properties outlined in Sec.~\ref{subsec:Definition-of-the}.
Therefore, the definition of the Berry phase Eq.~(\ref{eq:BerryPhase})
can be applied. It turns out that the standard CFL wave function has
a simple structure which makes possible an analytic determination
of the Berry phase.

To determine the Berry phase, we need to determine the matrix element
\begin{multline}
\Braket{\Psi_{\bm{k}}^{\mathrm{CF}}|e^{-\mathrm{i}\bm{q}\cdot\bm{z}_{1}}|\varphi_{\bm{k}^{\prime}}^{\mathrm{CF}}}=\Braket{\Psi_{\bm{k}}^{\mathrm{CF}}|\hat{P}_{\mathrm{LLL}}e^{-\mathrm{i}\bm{q}\cdot\bm{z}_{1}}|\varphi_{\bm{k}^{\prime}}^{\mathrm{CF}}}\\
=e^{-\frac{\left|q\right|^{2}}{4}}\Braket{\Psi_{\bm{k}}^{\mathrm{CF}}|\hat{t}_{1}\left(-\bm{q}\right)|\varphi_{\bm{k}^{\prime}}^{\mathrm{CF}}},\label{eq:matrixelement}
\end{multline}
where $\bm{k}^{\prime}\equiv\{\bm{k}_{1}+\bm{q},\bm{k}_{2},\dots\}$,
and we define the operator
\begin{equation}
\hat{t}_{i}\left(\bm{k}_{\alpha}\right)\equiv\exp\left(\mathrm{i}k_{\alpha}\partial_{z_{i}}+\frac{\mathrm{i}}{2}k_{\alpha}^{\ast}z_{i}\right).
\end{equation}
It is easy to verify the relation
\begin{equation}
\hat{t}_{i}\left(\bm{k}_{\alpha}\right)\hat{t}_{i}\left(\bm{k}_{\beta}\right)=e^{\frac{1}{2}\mathrm{i}\left(\bm{k}_{\alpha}\times\bm{k}_{\beta}\right)\cdot\bm{n}}\hat{t}_{i}\left(\bm{k}_{\alpha}+\bm{k}_{\beta}\right).\label{eq:tproduct}
\end{equation}

On the other hand, apart from an unimportant coefficient, the unsymmetrized
wave function $\varphi_{\bm{k}}^{\mathrm{CF}}$ can be written as%
\begin{equation}
\varphi_{\bm{k}^{\prime}}^{\mathrm{CF}}\left(z\right)=\hat{t}_{1}\left(\bm{k}_{1}+\bm{q}\right)\prod_{i\geq2}\hat{t}_{i}\left(\bm{k}_{i}\right)J\left(z\right).
\end{equation}
Using Eq.~(\ref{eq:tproduct}), we have
\begin{align}
\hat{t}_{1}\left(-\bm{q}\right)\varphi_{\bm{k}^{\prime}}^{\mathrm{CF}}\left(z\right) & =e^{-\frac{\textrm{i}}{2}(\bm{q}\times\bm{k}_{1})\cdot\bm{n}}\prod_{i}\hat{t}_{i}\left(\bm{k}_{i}\right)J\left(z\right)\nonumber \\
 & \equiv e^{-\frac{\textrm{i}}{2}(\bm{q}\times\bm{k}_{1})\cdot\bm{n}}\varphi_{\bm{k}}^{\mathrm{CF}}\left(z\right).
\end{align}
Inserting the relation into Eq.~(\ref{eq:matrixelement}), we obtain
\begin{equation}
\braket{\Psi_{\bm{k}}^{\mathrm{CF}}|e^{-\mathrm{i}\bm{q}\cdot\bm{z}_{1}}|\varphi_{\bm{k}^{\prime}}^{\mathrm{CF}}}=e^{\frac{1}{2}\textrm{i}(\bm{k}_{1}\times\bm{q})\cdot\bm{n}-\frac{1}{4}\vert q\vert^{2}}\braket{\Psi_{\bm{k}}^{\mathrm{CF}}|\Psi_{\bm{k}}^{\mathrm{CF}}}.
\end{equation}

Applying Eq.~(\ref{eq:BerryPhase}), we determine the Berry phase
in the e-representation:
\begin{equation}
\phi_{\bm{k}^{\prime}\bm{k}}^{\mathrm{e}}=\frac{1}{2}\left(\bm{q}\times\bm{k}_{1}\right)\cdot\bm{n}.
\end{equation}
The Berry phase in the CF (v-)representation can be determined by
applying Eq.~(\ref{eq:phiv}): 
\begin{equation}
\phi_{\bm{k}^{\prime}\bm{k}}^{\mathrm{v}}=-\frac{1}{2}\left(\bm{q}\times\bm{k}_{1}\right)\cdot\bm{n}.
\end{equation}

The Berry connections corresponding to the Berry phases are
\begin{equation}
\bm{A}_{\bm{k}_{1}}^{\mathrm{e/v}}=\pm(\bm{k}_{1}\times\bm{n})/2.
\end{equation}
They give rise to the Berry curvatures  (in the unit of $1/qB$)
\begin{equation}
\Omega_{\bm{k}_{1}}^{\mathrm{e/v}}\equiv(\bm{\nabla}_{\bm{k}_{1}}\times\bm{A}_{\bm{k}_{1}}^{\mathrm{e/v}})\cdot\bm{n}=\mp1.
\end{equation}
It indicates that the Berry curvature in the momentum space is a constant,
exactly the one suggested in the uniform-Berry-curvature picture of
CFs~\citep{shi2018,shi2017,haldane2016}.

\section{Berry phase of the JK wave function\label{sec:JK}}

In this section, we evaluate the Berry phase and Berry curvature 
of the JK wave function of the CFL. First, we introduce the JK wave
function on the torus geometry. Next, we describe the numerical implementation
of the calculations of the Berry phase. Then, we numerically evaluate
the Berry curvature and determine its distribution in the whole wave-vector
space.

\subsection{JK Wave function}

It is numerically hard to implement the LLL projection in Eq.\ (\ref{eq:CFL})
since it expands the wave function to $N!$ terms. To address the
issue, Jain and Kamilla introduce an alternative projection method~\citep{jain1997a}.
The projection method is adopted by Refs.~\citep{geraedts2018,wang_lattice_2019,wang_dirac_2019}
for numerically evaluating the Berry phase. The  wave function (JK
wave function) has the form~\citep{shao2015}
\begin{align}
\Psi_{\bm{k}}^{\mathrm{JK}}\left(z\right)= & \mathrm{det}\left[\psi_{i}(\bm{k}_{j})\right]\nonumber \\
 & \times\tilde{\sigma}^{m}\left(Z+\mathrm{i}K\right)\prod_{i<j}\tilde{\sigma}^{m-2}(z_{i}-z_{j}),\label{eq:JK}\\
\psi_{i}(\bm{k}_{j})= & e^{\mathrm{i}k_{j}^{\ast}z_{i}/2}\prod_{k\ne i}\tilde{\sigma}\left(z_{i}-z_{k}+\mathrm{i}mk_{j}-\mathrm{i}m\bar{k}\right),\label{eq:JK2}
\end{align}
with $\bar{k}\equiv K/N$. Its unsymmetrized form is
\begin{equation}
\varphi_{\bm{k}}^{\textrm{JK}}(z)=\tilde{\sigma}^{m}\left(Z+\mathrm{i}K\right)\prod_{i<j}\tilde{\sigma}^{m-2}(z_{i}-z_{j})\prod_{i}\psi_{i}(\bm{k}_{i}).\label{eq:JK3}
\end{equation}
The quantization of $\bm{k}$ is the same as that in Eq.~(\ref{eq:CFL}).
To evaluate the wave function, one only needs to calculate a determinant.
As a result, the computational complexity is greatly reduced compared
to the standard CFL wave function. This enables us to deal with large
systems numerically. 

It is easy to see that the JK wave function has the same center-of-mass
translational property as that of the standard CFL wave function (see
Sec.~\ref{subsec:Translational-symmetry}). Therefore, the Berry
phase definition Eq.~(\ref{eq:BerryPhase}) can be applied.

It is argued that the two wave functions are equivalent physically~\citep{jain1997a,jain2007}.
It turns out that this may not be true for evaluating the Berry phase,
as we will show next.

\subsection{Numerical implementation\label{subsec:Numerical-implementation}}

For both Geraedts et al.'s definition and our definition, the calculation
of the (Berry) phase involves the evaluation of a matrix element $\braket{\Psi_{\bm{k}}|\tilde{\Psi}_{\bm{k}^{\prime}}}$.
For Geraedts et al.'s definition, $\tilde{\Psi}_{\bm{k}^{\prime}}$
is defined by 
\begin{eqnarray}
\widetilde{\Psi}_{\bm{k'}} & = & e^{-\mathrm{i}\bm{q}_{1}\cdot\bm{z}_{1}}\Psi_{\bm{k}^{\prime}},
\end{eqnarray}
where we drop the superscript $\textrm{JK}$ for brevity. We note
that Geraedts et al.'s original definition uses the factor $\rho_{\bm{q}}=\sum_{i}e^{-\mathrm{i}\bm{q}\cdot\bm{z}_{i}}$.
The two forms are equivalent except  for an unimportant factor.

For our definition, after inserting $\hat{\mathcal{P}}$ into the
matrix element of Eq.~(\ref{eq:BerryPhase}), we can write $\widetilde{\Psi}_{\bm{k'}}$
as a form like  Eq.~(\ref{eq:JK})  but with the  determinant modified
to
\begin{equation}
\left|\begin{array}{cccc}
e^{-i\bm{q}_{1}\cdot\bm{z}_{1}}\psi_{1}(\bm{k}_{1}+\bm{q}_{1}) & \psi_{1}(\bm{k}_{2}) & \dots & \psi_{1}(\bm{k}_{N})\\
e^{-i\bm{q}_{1}\cdot\bm{z}_{2}}\psi_{2}(\bm{k}_{1}+\bm{q}_{1}) & \psi_{2}(\bm{k}_{2}) & \dots & \psi_{2}(\bm{k}_{N})\\
\vdots & \vdots & \ddots & \vdots\\
e^{-i\bm{q}_{1}\cdot\bm{z}_{N}}\psi_{N}(\bm{k}_{1}+\bm{q}_{1}) & \psi_{N}(\bm{k}_{2}) & \dots & \psi_{N}(\bm{k}_{N})
\end{array}\right|,
\end{equation}
i.e., the column with respect to the transported wave-vector ($\bm{k}_{1}$)
is modified by the ``momentum boost operator'' $e^{-\mathrm{i}\bm{q}_{1}\cdot\bm{z}_{i}}$.

We implement a Metropolis Monte-Carlo algorithm similar to that detailed
in Ref.~\citep{wang_lattice_2019}. Specifically, the overlap $|D|$
and phase $\phi$ of the matrix elements are evaluated by 
\begin{eqnarray}
|D|e^{i\phi} & \equiv & \frac{\langle\Psi_{\bm{k}}|\widetilde{\Psi}_{\bm{k'}}\rangle}{\sqrt{\langle\Psi_{\bm{k}}|\Psi_{\bm{k}}\rangle\langle\widetilde{\Psi}_{\bm{k}'}|\widetilde{\Psi}_{\bm{k}'}\rangle}}\nonumber \\
 & = & \frac{\frac{1}{\mathcal{N}}\sum'|\Psi_{\bm{k}}|^{2}\frac{\widetilde{\Psi}_{\bm{k}'}}{\Psi_{\bm{k}}}}{\sqrt{\frac{1}{\mathcal{N}}\sum'|\Psi_{\bm{k}}|^{2}\cdot\left|\frac{\widetilde{\Psi}_{\bm{k}'}}{\Psi_{\bm{k}}}\right|^{2}}},
\end{eqnarray}
where $\mathcal{N}$ denotes the normalization factor of $|\Psi_{\bm{k}}|^{2}$,
and $\sum^{'}$ stands for lattice summation of $z_{i}$'s~\citep{haldane2016,wang_lattice_2019}.
The Markov chains of our Monte-Carlo simulation sample $z_{i}$'s
by assuming a probability density $\propto|\Psi_{\bm{k}}(z)|^{2}$.
By using a Markov chain, we can determine the phases and overlaps
with respect to both the definitions simultaneously, since the two
definitions are only differed by $\tilde{\Psi}_{\bm{k}^{\prime}}$.

\subsection{Numerical results\label{subsec:Berry-curvature-distribution}}

To test our numerical implementation, we first calculate the Berry
phases along the paths calculated in Ref.~\citep{wang_lattice_2019}.
The numerical results are presented in Table.~\ref{table1}. For
Geraedts et al.'s definition, the results presented in Ref.~\citep{wang_lattice_2019}
and our own calculation results coincide well within numerical uncertainties.
The results with respect to our definition are also shown.

\begin{table}[tb]
\begin{tabular}{|c|c|l|c|c|}
\hline 
\multicolumn{2}{|c|}{Path} & \begin{tikzpicture}[scale=0.65,baseline=2em]
\draw[dashed,opacity=0.1][xstep=0.5,ystep=0.5] (-0.25,-0.25) grid (2.75,2.75);
\fill [blue] (canvas cs:x=0.5cm,y=1cm) circle (2pt);
\fill [blue] (canvas cs:x=0.5cm,y=1.5cm) circle (2pt);
\fill [blue] (canvas cs:x=1cm,y=0.5cm) circle (2pt);
\fill [blue] (canvas cs:x=1cm,y=1cm) circle (2pt);
\fill [blue] (canvas cs:x=1cm,y=1.5cm) circle (2pt);
\fill [blue] (canvas cs:x=1cm,y=2cm) circle (2pt);
\fill [blue] (canvas cs:x=1.5cm,y=0.5cm) circle (2pt);
\fill [blue] (canvas cs:x=1.5cm,y=1cm) circle (2pt);
\fill [blue] (canvas cs:x=1.5cm,y=1.5cm) circle (2pt);
\fill [blue] (canvas cs:x=1.5cm,y=2cm) circle (2pt);
\fill [blue] (canvas cs:x=2cm,y=1cm) circle (2pt);
\fill [blue] (canvas cs:x=2cm,y=1.5cm) circle (2pt);
\draw[ thick,draw=red]
(2.5,1) -- (2.5,1.5)-- (2,2)--(1.5,2.5)--(1,2.5)
--(0.5,2) --(0,1.5) --(0,1)--(0.5,0.5)--(1,0)
--(1.5,0) --(2,0.5) --(2.5,1);
\draw [->,red,thick] (0,1.5) to node[above] {} (0,1.15);
\node at ( 0.25,2.25)  {a)};
\end{tikzpicture} & \begin{tikzpicture}[scale=0.65,baseline=2em]
\draw[dashed,opacity=0.1][xstep=0.5,ystep=0.5] (-0.25,-0.25) grid (2.75,2.75);
\fill [blue] (canvas cs:x=0.5cm,y=1cm) circle (2pt);
\fill [blue] (canvas cs:x=0.5cm,y=1.5cm) circle (2pt);
\fill [blue] (canvas cs:x=1cm,y=0.5cm) circle (2pt);
\fill [blue] (canvas cs:x=1cm,y=1cm) circle (2pt);
\fill [blue] (canvas cs:x=1cm,y=1.5cm) circle (2pt);
\fill [blue] (canvas cs:x=1cm,y=2cm) circle (2pt);
\fill [blue] (canvas cs:x=1.5cm,y=0.5cm) circle (2pt);
\fill [blue] (canvas cs:x=1.5cm,y=1cm) circle (2pt);
\fill [blue] (canvas cs:x=1.5cm,y=1.5cm) circle (2pt);
\fill [blue] (canvas cs:x=1.5cm,y=2cm) circle (2pt);
\fill [blue] (canvas cs:x=2cm,y=1cm) circle (2pt);
\fill [blue] (canvas cs:x=2cm,y=1.5cm) circle (2pt);
\draw[ thick,draw=red]
(2,0.5) -- (2,2)--(0.5,2)--(0,1.5)--(0,1)--(0.5,0.5)--(2,0.5) ;
\draw [->,red,thick] (0,1.5) to node[above] {} (0,1.15);
\node at ( 0.25,2.25)  {b)};
\end{tikzpicture} & \begin{tikzpicture}[scale=0.65,baseline=2em]
\draw[dashed,opacity=0.1][xstep=0.5,ystep=0.5] (-0.25,-0.25) grid (2.75,2.75);
\fill [blue] (canvas cs:x=0.5cm,y=1cm) circle (2pt);
\fill [blue] (canvas cs:x=0.5cm,y=1.5cm) circle (2pt);
\fill [blue] (canvas cs:x=1cm,y=0.5cm) circle (2pt);
\fill [blue] (canvas cs:x=1cm,y=1cm) circle (2pt);
\fill [blue] (canvas cs:x=1cm,y=1.5cm) circle (2pt);
\fill [blue] (canvas cs:x=1cm,y=2cm) circle (2pt);
\fill [blue] (canvas cs:x=1.5cm,y=0.5cm) circle (2pt);
\fill [blue] (canvas cs:x=1.5cm,y=1cm) circle (2pt);
\fill [blue] (canvas cs:x=1.5cm,y=1.5cm) circle (2pt);
\fill [blue] (canvas cs:x=1.5cm,y=2cm) circle (2pt);
\fill [blue] (canvas cs:x=2cm,y=1cm) circle (2pt);
\fill [blue] (canvas cs:x=2cm,y=1.5cm) circle (2pt);
\draw[ thick,draw=red]
(0.5,2)--(0,1.5)--(0,1)--(0.5,0.5)--(0.5,2);
\draw [->,red,thick] (0,1.5) to node[above] {} (0,1.15);
\node at ( 0.25,2.25)  {c)};
\end{tikzpicture}\tabularnewline
\hline 
\hline 
\multirow{2}{*}{\begin{turn}{90}
old
\end{turn}} & Ref.~\citep{wang_lattice_2019} & 0.813(7) & 0.965(6) & -0.058(6)\tabularnewline
\cline{2-5} \cline{3-5} \cline{4-5} \cline{5-5} 
 & This work & 0.821(1) & 0.964(2) & -0.050(1)\tabularnewline
\hline 
\multicolumn{2}{|c|}{new} & 1.110 & 0.906 & 0.068(1)\tabularnewline
\hline 
\end{tabular}\caption{\label{table1} The phases for $N=13$. For Geraedts et al.'s definition
(old), both the results presented in Ref.~\citep{wang_lattice_2019}
and our own calculation are shown. The results with respect to our
definition (new) are shown in the last row. The numerical uncertainties
for paths a and b are not shown because they are too small.}
\end{table}

Berry phases with respect to both definitions for a few representative
paths are shown in Table.~\ref{table2}. An immediate observation
is that the calculation with our definition is much more robust numerically,
as evident from the magnitudes of the overlap. With our definition,
the overlap is always close to one and improves when $N$ is scaled
up. For Geraedts et al.'s definition, the overlap is nowhere close
to one and further deteriorates for larger $N$, and even nearly vanishes
for steps along directions perpendicular to the Fermi circle, resulting
in poor statistics and undeterminable results. Moreover, our definition
yields directly interpretable results, i.e., no subtraction of the
extraneous $\pm\pi/2$ phases noted in Ref.~\citep{geraedts2018}
is needed.

\begin{table}
\begin{tabular}{|c|c|c|c|c|c|c|c|}
\hline 
\multicolumn{2}{|c|}{Path} & \multicolumn{3}{c|}{\begin{tikzpicture}[scale=0.7,baseline=2em][<->]
\draw[dashed,opacity=0.1][xstep=0.25,ystep=0.25] (0,0) grid (2.5,2.5);
\fill [blue] (canvas cs:x=0.5cm,y=1cm) circle (1.25pt);
\fill [blue] (canvas cs:x=0.5cm,y=1.25cm) circle (1.25pt);
\fill [blue] (canvas cs:x=0.5cm,y=1.5cm) circle (1.25pt);
\fill [blue] (canvas cs:x=0.75cm,y=0.75cm) circle (1.25pt);
\fill [blue] (canvas cs:x=0.75cm,y=1cm) circle (1.25pt);
\fill [blue] (canvas cs:x=0.75cm,y=1.25cm) circle (1.25pt);
\fill [blue] (canvas cs:x=0.75cm,y=1.5cm) circle (1.25pt);
\fill [blue] (canvas cs:x=0.75cm,y=1.75cm) circle (1.25pt);
\fill [blue] (canvas cs:x=1cm,y=0.5cm) circle (1.25pt);
\fill [blue] (canvas cs:x=1cm,y=0.75cm) circle (1.25pt);
\fill [blue] (canvas cs:x=1cm,y=1cm) circle (1.25pt);
\fill [blue] (canvas cs:x=1cm,y=1.25cm) circle (1.25pt);
\fill [blue] (canvas cs:x=1cm,y=1.5cm) circle (1.25pt);
\fill [blue] (canvas cs:x=1cm,y=1.75cm) circle (1.25pt);
\fill [blue] (canvas cs:x=1cm,y=2cm) circle (1.25pt);
\fill [blue] (canvas cs:x=1.25cm,y=0.5cm) circle (1.25pt);
\fill [blue] (canvas cs:x=1.25cm,y=0.75cm) circle (1.25pt);
\fill [blue] (canvas cs:x=1.25cm,y=1cm) circle (1.25pt);
\fill [blue] (canvas cs:x=1.25cm,y=1.25cm) circle (1.25pt);
\fill [blue] (canvas cs:x=1.25cm,y=1.5cm) circle (1.25pt);
\fill [blue] (canvas cs:x=1.25cm,y=1.75cm) circle (1.25pt);
\fill [blue] (canvas cs:x=1.25cm,y=2cm) circle (1.25pt);
\fill [blue] (canvas cs:x=1.5cm,y=0.5cm) circle (1.25pt);
\fill [blue] (canvas cs:x=1.5cm,y=0.75cm) circle (1.25pt);
\fill [blue] (canvas cs:x=1.5cm,y=1cm) circle (1.25pt);
\fill [blue] (canvas cs:x=1.5cm,y=1.25cm) circle (1.25pt);
\fill [blue] (canvas cs:x=1.5cm,y=1.5cm) circle (1.25pt);
\fill [blue] (canvas cs:x=1.5cm,y=1.75cm) circle (1.25pt);
\fill [blue] (canvas cs:x=1.5cm,y=2cm) circle (1.25pt);
\fill [blue] (canvas cs:x=1.75cm,y=0.75cm) circle (1.25pt);
\fill [blue] (canvas cs:x=1.75cm,y=1cm) circle (1.25pt);
\fill [blue] (canvas cs:x=1.75cm,y=1.25cm) circle (1.25pt);
\fill [blue] (canvas cs:x=1.75cm,y=1.5cm) circle (1.25pt);
\fill [blue] (canvas cs:x=1.75cm,y=1.75cm) circle (1.25pt);
\fill [blue] (canvas cs:x=2cm,y=1cm) circle (1.25pt);
\fill [blue] (canvas cs:x=2cm,y=1.25cm) circle (1.25pt);
\fill [blue] (canvas cs:x=2cm,y=1.5cm) circle (1.25pt);
\draw[ thick,draw=red]
(2.25,1.25) -- (2.25,1.5)-- (2,1.75)--(1.75,2)--(1.5,2.25)
--(1.25,2.25) --(1,2.25) --(0.75,2)--(0.5,1.75)--(0.25,1.5)
--(0.25,1.25) --(0.25,1) --(0.5,0.75)--(0.75,0.5)--(1,0.25)
--(1.25,0.25) --(1.5,0.25) --(1.75,0.5)--(2,0.75)--(2.25,1)
--(2.25,1.25);
\draw [->,red,thick] (2.25,1) to node[right]  {}(2.25,1.25);
\node at ( 0.25,2.25)  {a)};
\end{tikzpicture}} & \multicolumn{2}{c|}{\begin{tikzpicture}[scale=0.7,baseline=2em][<->]
\draw[dashed,opacity=0.1][xstep=0.25,ystep=0.25] (0,0) grid (2.75,2.5);
\fill [blue] (canvas cs:x=0.5cm,y=1cm) circle (1.25pt);
\fill [blue] (canvas cs:x=0.5cm,y=1.25cm) circle (1.25pt);
\fill [blue] (canvas cs:x=0.5cm,y=1.5cm) circle (1.25pt);
\fill [blue] (canvas cs:x=0.75cm,y=0.75cm) circle (1.25pt);
\fill [blue] (canvas cs:x=0.75cm,y=1cm) circle (1.25pt);
\fill [blue] (canvas cs:x=0.75cm,y=1.25cm) circle (1.25pt);
\fill [blue] (canvas cs:x=0.75cm,y=1.5cm) circle (1.25pt);
\fill [blue] (canvas cs:x=0.75cm,y=1.75cm) circle (1.25pt);
\fill [blue] (canvas cs:x=1cm,y=0.5cm) circle (1.25pt);
\fill [blue] (canvas cs:x=1cm,y=0.75cm) circle (1.25pt);
\fill [blue] (canvas cs:x=1cm,y=1cm) circle (1.25pt);
\fill [blue] (canvas cs:x=1cm,y=1.25cm) circle (1.25pt);
\fill [blue] (canvas cs:x=1cm,y=1.5cm) circle (1.25pt);
\fill [blue] (canvas cs:x=1cm,y=1.75cm) circle (1.25pt);
\fill [blue] (canvas cs:x=1cm,y=2cm) circle (1.25pt);
\fill [blue] (canvas cs:x=1.25cm,y=0.5cm) circle (1.25pt);
\fill [blue] (canvas cs:x=1.25cm,y=0.75cm) circle (1.25pt);
\fill [blue] (canvas cs:x=1.25cm,y=1cm) circle (1.25pt);
\fill [blue] (canvas cs:x=1.25cm,y=1.25cm) circle (1.25pt);
\fill [blue] (canvas cs:x=1.25cm,y=1.5cm) circle (1.25pt);
\fill [blue] (canvas cs:x=1.25cm,y=1.75cm) circle (1.25pt);
\fill [blue] (canvas cs:x=1.25cm,y=2cm) circle (1.25pt);
\fill [blue] (canvas cs:x=1.5cm,y=0.5cm) circle (1.25pt);
\fill [blue] (canvas cs:x=1.5cm,y=0.75cm) circle (1.25pt);
\fill [blue] (canvas cs:x=1.5cm,y=1cm) circle (1.25pt);
\fill [blue] (canvas cs:x=1.5cm,y=1.25cm) circle (1.25pt);
\fill [blue] (canvas cs:x=1.5cm,y=1.5cm) circle (1.25pt);
\fill [blue] (canvas cs:x=1.5cm,y=1.75cm) circle (1.25pt);
\fill [blue] (canvas cs:x=1.5cm,y=2cm) circle (1.25pt);
\fill [blue] (canvas cs:x=1.75cm,y=0.75cm) circle (1.25pt);
\fill [blue] (canvas cs:x=1.75cm,y=1cm) circle (1.25pt);
\fill [blue] (canvas cs:x=1.75cm,y=1.25cm) circle (1.25pt);
\fill [blue] (canvas cs:x=1.75cm,y=1.5cm) circle (1.25pt);
\fill [blue] (canvas cs:x=1.75cm,y=1.75cm) circle (1.25pt);
\fill [blue] (canvas cs:x=2cm,y=1cm) circle (1.25pt);
\fill [blue] (canvas cs:x=2cm,y=1.25cm) circle (1.25pt);
\fill [blue] (canvas cs:x=2cm,y=1.5cm) circle (1.25pt);
\draw[ thick,dashed,draw=red,opacity=0.2]
(2.25,1.25) -- (2.25,1.5)-- (2,1.75)--(1.75,2)--(1.5,2.25)
--(1.25,2.25) --(1,2.25) --(0.75,2)--(0.5,1.75)--(0.25,1.5)
--(0.25,1.25) --(0.25,1) --(0.5,0.75)--(0.75,0.5)--(1,0.25)
--(1.25,0.25) --(1.5,0.25) --(1.75,0.5)--(2,0.75)--(2.25,1)
--(2.25,1.25);
\draw[ thick,draw=red](1.75,1.25)--(2,1.25)--(2,1.5)
--(1.75,1.5)--(1.75,1.25);
\draw [-,red,thick] (2,1.5) to node[above] {$1$} (1.75,1.5);
\draw[ thick,draw=red](2.25,1.25)--(2.5,1.25)--(2.5,1.5)
--(2.25,1.5)--(2.25,1.25);
\draw [-,red,thick] (2.5,1.5) to node[above] {$2$} (2.25,1.5);
\draw [->,red,thick] (2.5,1.5) to node[above] {} (2.3,1.5);
\draw [->,red,thick] (2,1.5) to node[above] {} (1.8,1.5);
\node at ( 0.25,2.25)  {b)};
\end{tikzpicture}} & \begin{tikzpicture}[scale=0.7,baseline=2em][<->]
\draw[dashed,opacity=0.1][xstep=0.25,ystep=0.25] (0,0) grid (2.5,2.5);
\fill [blue] (canvas cs:x=0.5cm,y=1cm) circle (1.25pt);
\fill [blue] (canvas cs:x=0.5cm,y=1.25cm) circle (1.25pt);
\fill [blue] (canvas cs:x=0.5cm,y=1.5cm) circle (1.25pt);
\fill [blue] (canvas cs:x=0.75cm,y=0.75cm) circle (1.25pt);
\fill [blue] (canvas cs:x=0.75cm,y=1cm) circle (1.25pt);
\fill [blue] (canvas cs:x=0.75cm,y=1.25cm) circle (1.25pt);
\fill [blue] (canvas cs:x=0.75cm,y=1.5cm) circle (1.25pt);
\fill [blue] (canvas cs:x=0.75cm,y=1.75cm) circle (1.25pt);
\fill [blue] (canvas cs:x=1cm,y=0.5cm) circle (1.25pt);
\fill [blue] (canvas cs:x=1cm,y=0.75cm) circle (1.25pt);
\fill [blue] (canvas cs:x=1cm,y=1cm) circle (1.25pt);
\fill [blue] (canvas cs:x=1cm,y=1.25cm) circle (1.25pt);
\fill [blue] (canvas cs:x=1cm,y=1.5cm) circle (1.25pt);
\fill [blue] (canvas cs:x=1cm,y=1.75cm) circle (1.25pt);
\fill [blue] (canvas cs:x=1cm,y=2cm) circle (1.25pt);
\fill [blue] (canvas cs:x=1.25cm,y=0.5cm) circle (1.25pt);
\fill [blue] (canvas cs:x=1.25cm,y=0.75cm) circle (1.25pt);
\fill [blue] (canvas cs:x=1.25cm,y=1cm) circle (1.25pt);
\fill [blue] (canvas cs:x=1.25cm,y=1.25cm) circle (1.25pt);
\fill [blue] (canvas cs:x=1.25cm,y=1.5cm) circle (1.25pt);
\fill [blue] (canvas cs:x=1.25cm,y=1.75cm) circle (1.25pt);
\fill [blue] (canvas cs:x=1.25cm,y=2cm) circle (1.25pt);
\fill [blue] (canvas cs:x=1.5cm,y=0.5cm) circle (1.25pt);
\fill [blue] (canvas cs:x=1.5cm,y=0.75cm) circle (1.25pt);
\fill [blue] (canvas cs:x=1.5cm,y=1cm) circle (1.25pt);
\fill [blue] (canvas cs:x=1.5cm,y=1.25cm) circle (1.25pt);
\fill [blue] (canvas cs:x=1.5cm,y=1.5cm) circle (1.25pt);
\fill [blue] (canvas cs:x=1.5cm,y=1.75cm) circle (1.25pt);
\fill [blue] (canvas cs:x=1.5cm,y=2cm) circle (1.25pt);
\fill [blue] (canvas cs:x=1.75cm,y=0.75cm) circle (1.25pt);
\fill [blue] (canvas cs:x=1.75cm,y=1cm) circle (1.25pt);
\fill [blue] (canvas cs:x=1.75cm,y=1.25cm) circle (1.25pt);
\fill [blue] (canvas cs:x=1.75cm,y=1.5cm) circle (1.25pt);
\fill [blue] (canvas cs:x=1.75cm,y=1.75cm) circle (1.25pt);
\fill [blue] (canvas cs:x=2cm,y=1cm) circle (1.25pt);
\fill [blue] (canvas cs:x=2cm,y=1.25cm) circle (1.25pt);
\fill [blue] (canvas cs:x=2cm,y=1.5cm) circle (1.25pt);
\draw[ thick,dashed,draw=red,opacity=0.2]
(2.25,1.25) -- (2.25,1.5)-- (2,1.75)--(1.75,2)--(1.5,2.25)
--(1.25,2.25) --(1,2.25) --(0.75,2)--(0.5,1.75)--(0.25,1.5)
--(0.25,1.25) --(0.25,1) --(0.5,0.75)--(0.75,0.5)--(1,0.25)
--(1.25,0.25) --(1.5,0.25) --(1.75,0.5)--(2,0.75)--(2.25,1)
--(2.25,1.25);
\draw[ thick,draw=red]
(1.75,1.25) -- (1.75,1.5)-- (1.5,1.75)--(1.25,1.75)--(1,1.75)
--(0.75,1.5) --(0.75,1.25) --(0.75,1)--(1,0.75)--(1.25,0.75)
--(1.5,0.75) --(1.75,1) --(1.75,1.25);
\draw [->,red,thick] (1.75,1) to node[above] {} (1.75,1.40);
\node at ( 0.25,2.25)  {c)};

\end{tikzpicture}\tabularnewline
\hline 
\multicolumn{2}{|c|}{$N$} & $13$ & $38$ & $110$ & $36$(b1) & $38$(b2) & $36$\tabularnewline
\hline 
\hline 
\multirow{2}{*}{$\phi^{\mathrm{v}}/\pi$} & old & $0.82$ & $0.72$ & $0.57$ & U.D. & U.D. & $0.93$\tabularnewline
\cline{2-8} \cline{3-8} \cline{4-8} \cline{5-8} \cline{6-8} \cline{7-8} \cline{8-8} 
 & new & $1.11$ & $1.03$ & $1.01$ & $0.75^{\ast}$ & $0.05^{\ast}$ & $0.61$\tabularnewline
\hline 
\hline 
\multirow{2}{*}{$|D|_{\mathrm{min}}$} & old & $0.65^{\ast}$ & $0.36^{\ast}$ & $0.18^{\ast}$ & $0.04^{\ast}$ & $0.01^{\ast}$ & $0.22^{\ast}$\tabularnewline
\cline{2-8} \cline{3-8} \cline{4-8} \cline{5-8} \cline{6-8} \cline{7-8} \cline{8-8} 
 & new & $0.94$ & $0.98$ & $0.99$ & $0.99$ & $0.99$ & $0.99$\tabularnewline
\hline 
\end{tabular}\caption{\label{table2} CF Berry phases $\phi^{\mathrm{v}}$ and minimal overlaps
$|D|_{\mathrm{min}}$ along different paths for the JK wave function.
The paths are indicated by arrowed solid lines comprised of steps
with minimal changes of the quantized wave vectors. Three kinds of
paths are considered: a) the Fermi circle; b) a unit plaquette inside
(b1) or outside (b2) the Fermi sea; (c) a closed path inside the Fermi
sea. Both results for our definition (new) and Geraedts et al.'s definition
(old) are shown. $|D|_{\mathrm{min}}$ is the minimum overlap among
steps along a path. For the paths inside the Fermi sea, a hole is
transported, and resulting Berry phases are shown with inverted signs.
The values marked with $\ast$ have been scaled by a factor of $N$.
U.D. indicates an undeterminable result due to a vanishing overlap.}
\end{table}

It is interesting to observe that the two different definitions actually
lead to similar qualitative conclusions as long as $\phi^{\mathrm{v}}$
is interpreted as the CF Berry phase. With our definition, the Berry
phase of adiabatic transport of a CF around the Fermi circle is converged
to $\pi$ (path a, $N=110$), whereas with Geraedts et al.'s definition,
it involves guesswork to reach the same conclusion. We also find that
the Berry phase for transport around a unit plaquette outside the
Fermi sea (path b2) nearly vanishes. This is consistent with Geraedts
et al.'s observation that the phase is independent of the area of
the trajectory enclosing the Fermi sea. The consistencies may not
be a coincidence.  See Sec.~\ref{subsec:Interpretation-of-Geraedts}
for an interpretation.

\begin{figure}[t]
\includegraphics[width=1\columnwidth]{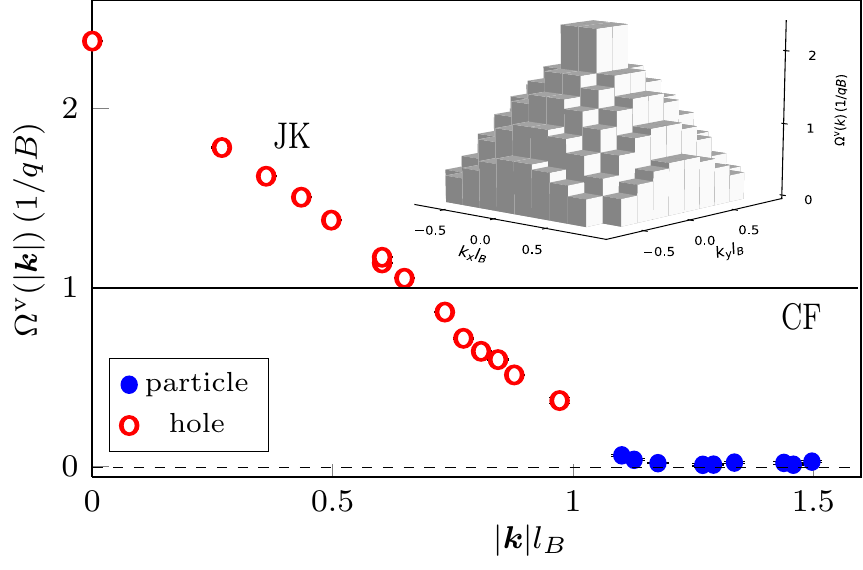}

\caption{\label{figure1} The Berry curvature $\Omega^{\mathrm{v}}(|\bm{k}|)$
as a function of the CF wave number $|\bm{k}|$ for the half-filled
CFL ($m=2$). The Berry curvature for the JK wave function is numerically
determined by transporting a CF (hole) outside (inside) the Fermi
sea consist of 109 CFs, shown as filled (empty) dots. The inset bar
plot shows its distribution on the 2D plane of the momentum space.
The Berry curvature for the standard CFL wave function is equal to
one, shown as the solid line.}
\end{figure}

The distribution of the Berry curvature, both inside and outside the
Fermi sea, can now be determined because of the improved numerical
robustness. To determine the Berry curvature, we transport a CF or
a hole along the edges of a unit plaquette (see path b in Table.~\ref{table2}),
and the Berry curvature for the plaquette is determined by $\Omega^{\mathrm{v}}=\phi_{\mathrm{B}}^{\mathrm{v}}/S_{0}$,
where $S_{0}=2\pi/N_{\phi}$ is the area of the unit plaquette. The
result is shown in Fig.~\ref{figure1}. We see that the Berry curvature
has a continuous distribution inside the Fermi sea and vanishes outside.
The distribution is obviously \emph{not} the singular one implied
by the Dirac interpretation~\citep{wang_dirac_2019,goldman2018a}.

\section{Uniform background of the Berry curvature\label{sec:Uniform-background}}

It is evident that for both the wave functions, the Berry curvature
is a constant in most of the region of the $\bm{k}$-space except
the one occupied by  the Fermi sea. In other words, the Berry curvature
has a uniform background. The values of the background for the two
wave functions are different: $1$ for the standard CFL wave function
and $0$ for the JK wave function, respectively.   In this section,
we show that the values   can be  determined analytically by inspecting
the quasi-periodicities of the wave functions in the $\bm{k}$-space~\citep{thouless1984}.
It turns out that the two wave functions have different quasi-periodicities,
 resulting in the different values of the uniform background.

We first examine the standard CFL wave function Eq.~(\ref{eq:CFL2}).
By using Eq.~(\ref{eq:sigma}), it is easy to show that it has the
 quasiperiodicity in the $\bm{k}$-space:
\begin{align}
\left.\Psi_{\bm{k}}^{\mathrm{CF}}\right|_{\bm{k}_{1}\rightarrow\bm{k}_{1}+\bm{L}\times\bm{n}}= & \xi^{N_{\phi}}(L)e^{\frac{L^{\ast}}{2}\left(\mathrm{i}k_{1}+\frac{1}{2}L\right)}\Psi_{\bm{k}}^{\mathrm{CF}},\label{eq:kP}
\end{align}
where $\bm{k}_{1}\rightarrow\bm{k}_{1}+\bm{L}\times\bm{n}$ corresponds
to $\mathrm{i}k_{1}\rightarrow\mathrm{i}k_{1}+L$ in the complex form.

As a result, we can define a super-Brillouin zone (SBZ) spanned by
$\bm{K}_{\alpha}=\bm{L}_{\alpha}\times\bm{n}$, $\alpha=1,2$ with
$\bm{L}_{1}$ and $\bm{L}_{2}$ being the two edges of the torus.
From Eq.~(\ref{eq:kP}), by applying either the original definition
Eq.~(\ref{eq:gammaC})~\footnote{Note that for unnormalized wave functions, the Berry connection is
determined by the formula $\bm{A}_{\bm{\alpha}}=-\mathrm{Im}\braket{\Psi_{\bm{\alpha}}|\partial_{\bm{\alpha}}\Psi_{\bm{\alpha}}}/\braket{\Psi_{\bm{\alpha}}|\Psi_{\bm{\alpha}}}$.} or the generalized definition Eq.~(\ref{eq:Ak1}), we find that
the Berry connection has the quasi-periodicity:
\begin{equation}
\bm{A}_{\bm{k}_{1}+\bm{K}}^{\mathrm{e}}=\bm{A}_{\bm{k}_{1}}^{\mathrm{e}}+\frac{1}{2}(\bm{K}\times\bm{n}),\label{eq:Aperiod}
\end{equation}
where $\bm{K}$ is one of reciprocal lattice vectors of the SBZ. The
total Chern number of the SBZ can be determined by $C_{\mathrm{tot}}=(2\pi)^{-1}\oint\bm{A}_{\bm{k}_{1}}^{\mathrm{e}}\cdot\mathrm{d}\bm{k}_{1}$
with the circuit integral along the boundary of the SBZ~\citep{thouless1984}.
Using Eq.~(\ref{eq:Aperiod}), it is easy to show that the integral
is equal to $-A/2\pi=-N_{\phi}$. Since the Berry curvature is a constant
in most of region of the SBZ, $C_{\mathrm{tot}}$ is equal to $\bar{\Omega}^{\mathrm{e}}A/2\pi$
in the limit of $A\rightarrow\infty$, where $\bar{\Omega}^{\mathrm{e}}$
is the value of the uniform background of the Berry curvature.  We
thus obtain
\begin{equation}
\bar{\Omega}^{\mathrm{e}}=-1,
\end{equation}
for the standard CFL wave function. For the CF representation, we
have
\begin{equation}
\bar{\Omega}^{\mathrm{v}}=1.
\end{equation}

Similarly, it is easy to show that $\Psi_{\bm{k}}^{\mathrm{JK}}$
has an approximated quasi-periodicity: 
\begin{equation}
\left.\Psi_{\bm{k}}^{\mathrm{JK}}\right|_{\bm{k}_{1}\rightarrow\bm{k}_{1}+\bm{L}\times\bm{n}}\propto\exp(\mathrm{i}mL^{\ast}k_{1}/2)\Psi_{\bm{k}}^{\mathrm{JK}},
\end{equation}
where we ignore the small change of $\bar{k}$ in the limit of $N\rightarrow\infty$.
The presence of $m$ in the phase factor is notable. It originates
from the $\mathrm{i}mk_{j}$ factor in the argument of $\tilde{\sigma}$-function
in Eq.~(\ref{eq:JK2})~\citep{shao2015,pu2017}. The quasi-periodicity
of the Berry connection is modified to
\begin{equation}
\bm{A}_{\bm{k}_{1}+\bm{K}}^{\mathrm{e}}=\bm{A}_{\bm{k}_{1}}^{\mathrm{e}}+\frac{m}{2}(\bm{K}\times\bm{n}).\label{eq:Aperiod-1}
\end{equation}
By applying the same analysis as that for the standard CFL wave function,
 we obtain that the uniform background  of the Berry curvature for
the JK wave function is $-m$ in the e-representation and $2-m$ in
the CF representation.

We summarize  the values of the Berry curvature background as follows:%
\begin{equation}
\bar{\Omega}^{\mathrm{v}}=\begin{cases}
1, & \mathrm{(CF)}\\
2-m, & \mathrm{(JK)}
\end{cases}.
\end{equation}
Note that the result is  solely determined by the quasi-periodicities
 of the wave functions.  The fact that the two wave functions have
different quasi-periodicities means that they must have different
Berry curvatures.

\section{Discussion and Summary\label{sec:Summary}}

In summary, we have (a) derived the definition of the Berry phase
applicable for CFL systems; (b) analytically determined the Berry
phase of the CFL with respect to the standard CF wave function, and
found that it yields a uniform Berry curvature; (c) numerically calculated
the Berry phase with respect to the JK wave function, and determined
the distribution of the Berry curvature in the whole momentum space;
(d) analytically shown that the Berry phases with respect to the two
wave functions must be different because of their different quasi-periodicities. 

For both the wave functions, we find that a CF adiabatically transported
around the Fermi circle acquires a Berry phase $\pi$ in the CF representation.
Since the Berry phase can be interpreted as the intrinsic anomalous
Hall conductance~\citep{haldane2004} (in the unit of $-e^{2}/2\pi h$
for $\sigma_{xy}$~\citep{jungwirth2002}), both the wave functions
can correctly predict the Hall conductance of CFs for a particle-hole
symmetric half-filled Landau level~\citep{kivelson1997}, in both
its magnitude and sign. The result is actually consistent with the
Dirac theory. However, microscopically, both the Berry curvature distributions
with respect to  the two wave functions are \emph{not} the singular
one implied by the  Dirac picture. We thus expect that the effective
theories with respect to the two wave functions are not the Dirac
theory when physics away from the Fermi level is concerned.%

On the other hand, one may question the physical relevance of these
subtle differences between different effective theories. Indeed, up
to now, most of predictions of different effective theories are focused
on the effects of the $\pi$-Berry phase, and indistinguishable. It
doesn't help that the HLR theory, which has no $\pi$-Berry phase,
can also correctly predict the Hall conductance of CFL by considering
the effect of scattering by the fluctuation of the effective magnetic
field~\citep{wang_particle-hole_2017}.  The situation is actually
typical, as for other theories in a similar stage when different pictures
compete and seem to provide equally good explanations for a limited
set of observations. For the CF theory, we would like to argue that:
(a) a wave function must have one and only one correct effective theory,
unless different effective theories can be shown equivalent; (b) microscopic
details of different effective theories are relevant because they
may lead to different physical predictions. One such example is shown
in Ref.~\citep{ji_asymmetry_2020}, which indicates that different
ways of modulating CF systems can induce different asymmetries in
geometric resonance experiments~\citep{kamburov2014} as a result
of the ``subatomic'' dipole structure of the CF. Predictions like
that could be tested experimentally and differentiate different effective
theories; (c) correct microscopic details may be the key of obtaining
a consistent effective theory free of difficulties such as the effective
mass divergence~\citep{simon1998}.

\begin{acknowledgments}
J.S. thanks F. D. M. Haldane for sharing the link to Ref.~\citep{haldane2016},
and thanks F. D. M. Haldane and Jie Wang for valuable discussions.
This work is supported by National Basic Research Program of China
(973 Program) Grant No. 2015CB921101 and National Science Foundation
of China Grant No. 11325416. 
\end{acknowledgments}

\bibliographystyle{apsrev4-1}
\bibliography{BerryCFL}

\begin{thebibliography}{37}%
\makeatletter
\providecommand \@ifxundefined [1]{%
 \@ifx{#1\undefined}
}%
\providecommand \@ifnum [1]{%
 \ifnum #1\expandafter \@firstoftwo
 \else \expandafter \@secondoftwo
 \fi
}%
\providecommand \@ifx [1]{%
 \ifx #1\expandafter \@firstoftwo
 \else \expandafter \@secondoftwo
 \fi
}%
\providecommand \natexlab [1]{#1}%
\providecommand \enquote  [1]{``#1''}%
\providecommand \bibnamefont  [1]{#1}%
\providecommand \bibfnamefont [1]{#1}%
\providecommand \citenamefont [1]{#1}%
\providecommand \href@noop [0]{\@secondoftwo}%
\providecommand \href [0]{\begingroup \@sanitize@url \@href}%
\providecommand \@href[1]{\@@startlink{#1}\@@href}%
\providecommand \@@href[1]{\endgroup#1\@@endlink}%
\providecommand \@sanitize@url [0]{\catcode `\\12\catcode `\$12\catcode
  `\&12\catcode `\#12\catcode `\^12\catcode `\_12\catcode `\%12\relax}%
\providecommand \@@startlink[1]{}%
\providecommand \@@endlink[0]{}%
\providecommand \url  [0]{\begingroup\@sanitize@url \@url }%
\providecommand \@url [1]{\endgroup\@href {#1}{\urlprefix }}%
\providecommand \urlprefix  [0]{URL }%
\providecommand \Eprint [0]{\href }%
\providecommand \doibase [0]{http://dx.doi.org/}%
\providecommand \selectlanguage [0]{\@gobble}%
\providecommand \bibinfo  [0]{\@secondoftwo}%
\providecommand \bibfield  [0]{\@secondoftwo}%
\providecommand \translation [1]{[#1]}%
\providecommand \BibitemOpen [0]{}%
\providecommand \bibitemStop [0]{}%
\providecommand \bibitemNoStop [0]{.\EOS\space}%
\providecommand \EOS [0]{\spacefactor3000\relax}%
\providecommand \BibitemShut  [1]{\csname bibitem#1\endcsname}%
\let\auto@bib@innerbib\@empty
\bibitem [{\citenamefont {Xiao}\ \emph {et~al.}(2010)\citenamefont {Xiao},
  \citenamefont {Chang},\ and\ \citenamefont {Niu}}]{xiao2010}%
  \BibitemOpen
  \bibfield  {author} {\bibinfo {author} {\bibfnamefont {D.}~\bibnamefont
  {Xiao}}, \bibinfo {author} {\bibfnamefont {M.-C.}\ \bibnamefont {Chang}}, \
  and\ \bibinfo {author} {\bibfnamefont {Q.}~\bibnamefont {Niu}},\ }\href
  {\doibase 10.1103/RevModPhys.82.1959} {\bibfield  {journal} {\bibinfo
  {journal} {Rev. Mod. Phys.}\ }\textbf {\bibinfo {volume} {82}},\ \bibinfo
  {pages} {1959} (\bibinfo {year} {2010})}\BibitemShut {NoStop}%
\bibitem [{\citenamefont {Hasan}\ and\ \citenamefont {Kane}(2010)}]{hasan2010}%
  \BibitemOpen
  \bibfield  {author} {\bibinfo {author} {\bibfnamefont {M.~Z.}\ \bibnamefont
  {Hasan}}\ and\ \bibinfo {author} {\bibfnamefont {C.~L.}\ \bibnamefont
  {Kane}},\ }\href {\doibase 10.1103/RevModPhys.82.3045} {\bibfield  {journal}
  {\bibinfo  {journal} {Rev. Mod. Phys.}\ }\textbf {\bibinfo {volume} {82}},\
  \bibinfo {pages} {3045} (\bibinfo {year} {2010})}\BibitemShut {NoStop}%
\bibitem [{\citenamefont {Jia}\ \emph {et~al.}(2016)\citenamefont {Jia},
  \citenamefont {Xu},\ and\ \citenamefont {Hasan}}]{jia2016}%
  \BibitemOpen
  \bibfield  {author} {\bibinfo {author} {\bibfnamefont {S.}~\bibnamefont
  {Jia}}, \bibinfo {author} {\bibfnamefont {S.-Y.}\ \bibnamefont {Xu}}, \ and\
  \bibinfo {author} {\bibfnamefont {M.~Z.}\ \bibnamefont {Hasan}},\ }\href
  {\doibase 10.1038/nmat4787} {\bibfield  {journal} {\bibinfo  {journal} {Nat
  Mater}\ }\textbf {\bibinfo {volume} {15}},\ \bibinfo {pages} {1140} (\bibinfo
  {year} {2016})}\BibitemShut {NoStop}%
\bibitem [{\citenamefont {Cao}\ \emph {et~al.}(2012)\citenamefont {Cao},
  \citenamefont {Wang}, \citenamefont {Han}, \citenamefont {Ye}, \citenamefont
  {Zhu}, \citenamefont {Shi}, \citenamefont {Niu}, \citenamefont {Tan},
  \citenamefont {Wang}, \citenamefont {Liu},\ and\ \citenamefont
  {Feng}}]{cao2012}%
  \BibitemOpen
  \bibfield  {author} {\bibinfo {author} {\bibfnamefont {T.}~\bibnamefont
  {Cao}}, \bibinfo {author} {\bibfnamefont {G.}~\bibnamefont {Wang}}, \bibinfo
  {author} {\bibfnamefont {W.}~\bibnamefont {Han}}, \bibinfo {author}
  {\bibfnamefont {H.}~\bibnamefont {Ye}}, \bibinfo {author} {\bibfnamefont
  {C.}~\bibnamefont {Zhu}}, \bibinfo {author} {\bibfnamefont {J.}~\bibnamefont
  {Shi}}, \bibinfo {author} {\bibfnamefont {Q.}~\bibnamefont {Niu}}, \bibinfo
  {author} {\bibfnamefont {P.}~\bibnamefont {Tan}}, \bibinfo {author}
  {\bibfnamefont {E.}~\bibnamefont {Wang}}, \bibinfo {author} {\bibfnamefont
  {B.}~\bibnamefont {Liu}}, \ and\ \bibinfo {author} {\bibfnamefont
  {J.}~\bibnamefont {Feng}},\ }\href {\doibase 10.1038/ncomms1882} {\bibfield
  {journal} {\bibinfo  {journal} {Nature Communications}\ }\textbf {\bibinfo
  {volume} {3}},\ \bibinfo {pages} {887} (\bibinfo {year} {2012})}\BibitemShut
  {NoStop}%
\bibitem [{\citenamefont {Jain}(2007)}]{jain2007}%
  \BibitemOpen
  \bibfield  {author} {\bibinfo {author} {\bibfnamefont {J.~K.}\ \bibnamefont
  {Jain}},\ }\href@noop {} {{\selectlanguage {English}\emph {\bibinfo {title}
  {Composite Fermions}}}}\ (\bibinfo  {publisher} {{Cambridge University
  Press}},\ \bibinfo {year} {2007})\BibitemShut {NoStop}%
\bibitem [{\citenamefont {Jain}\ and\ \citenamefont
  {Anderson}(2009)}]{jain2009}%
  \BibitemOpen
  \bibfield  {author} {\bibinfo {author} {\bibfnamefont {J.~K.}\ \bibnamefont
  {Jain}}\ and\ \bibinfo {author} {\bibfnamefont {P.~W.}\ \bibnamefont
  {Anderson}},\ }\href {\doibase 10.1073/pnas.0902901106} {\bibfield  {journal}
  {\bibinfo  {journal} {Proceedings of the National Academy of Sciences}\
  }\textbf {\bibinfo {volume} {106}},\ \bibinfo {pages} {9131} (\bibinfo {year}
  {2009})}\BibitemShut {NoStop}%
\bibitem [{\citenamefont {Balram}\ and\ \citenamefont
  {Jain}(2016)}]{balram_nature_2016}%
  \BibitemOpen
  \bibfield  {author} {\bibinfo {author} {\bibfnamefont {A.~C.}\ \bibnamefont
  {Balram}}\ and\ \bibinfo {author} {\bibfnamefont {J.~K.}\ \bibnamefont
  {Jain}},\ }\href {\doibase 10.1103/PhysRevB.93.235152} {\bibfield  {journal}
  {\bibinfo  {journal} {Phys. Rev. B}\ }\textbf {\bibinfo {volume} {93}},\
  \bibinfo {pages} {235152} (\bibinfo {year} {2016})}\BibitemShut {NoStop}%
\bibitem [{\citenamefont {Kalmeyer}\ and\ \citenamefont
  {Zhang}(1992)}]{kalmeyer1992}%
  \BibitemOpen
  \bibfield  {author} {\bibinfo {author} {\bibfnamefont {V.}~\bibnamefont
  {Kalmeyer}}\ and\ \bibinfo {author} {\bibfnamefont {S.-C.}\ \bibnamefont
  {Zhang}},\ }\href {\doibase 10.1103/PhysRevB.46.9889} {\bibfield  {journal}
  {\bibinfo  {journal} {Phys. Rev. B}\ }\textbf {\bibinfo {volume} {46}},\
  \bibinfo {pages} {9889} (\bibinfo {year} {1992})}\BibitemShut {NoStop}%
\bibitem [{\citenamefont {Halperin}\ \emph {et~al.}(1993)\citenamefont
  {Halperin}, \citenamefont {Lee},\ and\ \citenamefont {Read}}]{halperin1993}%
  \BibitemOpen
  \bibfield  {author} {\bibinfo {author} {\bibfnamefont {B.~I.}\ \bibnamefont
  {Halperin}}, \bibinfo {author} {\bibfnamefont {P.~A.}\ \bibnamefont {Lee}}, \
  and\ \bibinfo {author} {\bibfnamefont {N.}~\bibnamefont {Read}},\ }\href
  {\doibase 10.1103/PhysRevB.47.7312} {\bibfield  {journal} {\bibinfo
  {journal} {Phys. Rev. B}\ }\textbf {\bibinfo {volume} {47}},\ \bibinfo
  {pages} {7312} (\bibinfo {year} {1993})}\BibitemShut {NoStop}%
\bibitem [{\citenamefont {Simon}(1998)}]{simon1998}%
  \BibitemOpen
  \bibfield  {author} {\bibinfo {author} {\bibfnamefont {S.~H.}\ \bibnamefont
  {Simon}},\ }in\ \href@noop {} {\emph {\bibinfo {booktitle} {Composite
  {{Fermions}}}}},\ \bibinfo {editor} {edited by\ \bibinfo {editor}
  {\bibfnamefont {O.}~\bibnamefont {Heinonen}}}\ (\bibinfo  {publisher} {{World
  Scientific}},\ \bibinfo {year} {1998})\ p.~\bibinfo {pages} {91}\BibitemShut
  {NoStop}%
\bibitem [{\citenamefont {Kivelson}\ \emph {et~al.}(1997)\citenamefont
  {Kivelson}, \citenamefont {Lee}, \citenamefont {Krotov},\ and\ \citenamefont
  {Gan}}]{kivelson1997}%
  \BibitemOpen
  \bibfield  {author} {\bibinfo {author} {\bibfnamefont {S.~A.}\ \bibnamefont
  {Kivelson}}, \bibinfo {author} {\bibfnamefont {D.-H.}\ \bibnamefont {Lee}},
  \bibinfo {author} {\bibfnamefont {Y.}~\bibnamefont {Krotov}}, \ and\ \bibinfo
  {author} {\bibfnamefont {J.}~\bibnamefont {Gan}},\ }\href {\doibase
  10.1103/PhysRevB.55.15552} {\bibfield  {journal} {\bibinfo  {journal} {Phys.
  Rev. B}\ }\textbf {\bibinfo {volume} {55}},\ \bibinfo {pages} {15552}
  (\bibinfo {year} {1997})}\BibitemShut {NoStop}%
\bibitem [{\citenamefont {Son}(2015)}]{son2015}%
  \BibitemOpen
  \bibfield  {author} {\bibinfo {author} {\bibfnamefont {D.~T.}\ \bibnamefont
  {Son}},\ }\href {\doibase 10.1103/PhysRevX.5.031027} {\bibfield  {journal}
  {\bibinfo  {journal} {Phys. Rev. X}\ }\textbf {\bibinfo {volume} {5}},\
  \bibinfo {pages} {031027} (\bibinfo {year} {2015})}\BibitemShut {NoStop}%
\bibitem [{\citenamefont {Sundaram}\ and\ \citenamefont
  {Niu}(1999)}]{sundaram1999}%
  \BibitemOpen
  \bibfield  {author} {\bibinfo {author} {\bibfnamefont {G.}~\bibnamefont
  {Sundaram}}\ and\ \bibinfo {author} {\bibfnamefont {Q.}~\bibnamefont {Niu}},\
  }\href {\doibase 10.1103/PhysRevB.59.14915} {\bibfield  {journal} {\bibinfo
  {journal} {Phys. Rev. B}\ }\textbf {\bibinfo {volume} {59}},\ \bibinfo
  {pages} {14915} (\bibinfo {year} {1999})}\BibitemShut {NoStop}%
\bibitem [{\citenamefont {Haldane}(2016)}]{haldane2016}%
  \BibitemOpen
  \bibfield  {author} {\bibinfo {author} {\bibfnamefont {F.~D.~M.}\
  \bibnamefont {Haldane}},\ }\href@noop {} {\enquote {\bibinfo {title} {A model
  wavefunction for the composite {{Fermi}} liquid: Its geometry and
  entanglement},}\ } (\bibinfo {year} {2016}),\ \bibinfo {note} {in APS March
  Meeting, Baltimore}\BibitemShut {NoStop}%
\bibitem [{\citenamefont {Shi}()}]{shi2017}%
  \BibitemOpen
  \bibfield  {author} {\bibinfo {author} {\bibfnamefont {J.}~\bibnamefont
  {Shi}},\ }\href@noop {} {}\Eprint {http://arxiv.org/abs/1704.07712}
  {arXiv:1704.07712} \BibitemShut {NoStop}%
\bibitem [{\citenamefont {Shi}\ and\ \citenamefont {Ji}(2018)}]{shi2018}%
  \BibitemOpen
  \bibfield  {author} {\bibinfo {author} {\bibfnamefont {J.}~\bibnamefont
  {Shi}}\ and\ \bibinfo {author} {\bibfnamefont {W.}~\bibnamefont {Ji}},\
  }\href {\doibase 10.1103/PhysRevB.97.125133} {\bibfield  {journal} {\bibinfo
  {journal} {Phys. Rev. B}\ }\textbf {\bibinfo {volume} {97}},\ \bibinfo
  {pages} {125133} (\bibinfo {year} {2018})}\BibitemShut {NoStop}%
\bibitem [{\citenamefont {Read}(1994)}]{read1994}%
  \BibitemOpen
  \bibfield  {author} {\bibinfo {author} {\bibfnamefont {N.}~\bibnamefont
  {Read}},\ }\href {\doibase 10.1088/0268-1242/9/11S/002} {\bibfield  {journal}
  {\bibinfo  {journal} {Semicond. Sci. Technol.}\ }\textbf {\bibinfo {volume}
  {9}},\ \bibinfo {pages} {1859} (\bibinfo {year} {1994})}\BibitemShut
  {NoStop}%
\bibitem [{\citenamefont {Geraedts}\ \emph {et~al.}(2018)\citenamefont
  {Geraedts}, \citenamefont {Wang}, \citenamefont {Rezayi},\ and\ \citenamefont
  {Haldane}}]{geraedts2018}%
  \BibitemOpen
  \bibfield  {author} {\bibinfo {author} {\bibfnamefont {S.~D.}\ \bibnamefont
  {Geraedts}}, \bibinfo {author} {\bibfnamefont {J.}~\bibnamefont {Wang}},
  \bibinfo {author} {\bibfnamefont {E.~H.}\ \bibnamefont {Rezayi}}, \ and\
  \bibinfo {author} {\bibfnamefont {F.~D.~M.}\ \bibnamefont {Haldane}},\ }\href
  {\doibase 10.1103/PhysRevLett.121.147202} {\bibfield  {journal} {\bibinfo
  {journal} {Phys. Rev. Lett.}\ }\textbf {\bibinfo {volume} {121}},\ \bibinfo
  {pages} {147202} (\bibinfo {year} {2018})}\BibitemShut {NoStop}%
\bibitem [{\citenamefont {Wang}\ \emph {et~al.}(2019)\citenamefont {Wang},
  \citenamefont {Geraedts}, \citenamefont {Rezayi},\ and\ \citenamefont
  {Haldane}}]{wang_lattice_2019}%
  \BibitemOpen
  \bibfield  {author} {\bibinfo {author} {\bibfnamefont {J.}~\bibnamefont
  {Wang}}, \bibinfo {author} {\bibfnamefont {S.~D.}\ \bibnamefont {Geraedts}},
  \bibinfo {author} {\bibfnamefont {E.~H.}\ \bibnamefont {Rezayi}}, \ and\
  \bibinfo {author} {\bibfnamefont {F.~D.~M.}\ \bibnamefont {Haldane}},\ }\href
  {\doibase 10.1103/PhysRevB.99.125123} {\bibfield  {journal} {\bibinfo
  {journal} {Phys. Rev. B}\ }\textbf {\bibinfo {volume} {99}},\ \bibinfo
  {pages} {125123} (\bibinfo {year} {2019})}\BibitemShut {NoStop}%
\bibitem [{\citenamefont {Wang}(2019)}]{wang_dirac_2019}%
  \BibitemOpen
  \bibfield  {author} {\bibinfo {author} {\bibfnamefont {J.}~\bibnamefont
  {Wang}},\ }\href {\doibase 10.1103/PhysRevLett.122.257203} {\bibfield
  {journal} {\bibinfo  {journal} {Phys. Rev. Lett.}\ }\textbf {\bibinfo
  {volume} {122}},\ \bibinfo {pages} {257203} (\bibinfo {year}
  {2019})}\BibitemShut {NoStop}%
\bibitem [{\citenamefont {Jain}\ and\ \citenamefont
  {Kamilla}(1997)}]{jain1997a}%
  \BibitemOpen
  \bibfield  {author} {\bibinfo {author} {\bibfnamefont {J.~K.}\ \bibnamefont
  {Jain}}\ and\ \bibinfo {author} {\bibfnamefont {R.~K.}\ \bibnamefont
  {Kamilla}},\ }\href {\doibase 10.1142/S0217979297001301} {\bibfield
  {journal} {\bibinfo  {journal} {Int. J. Mod. Phys. B}\ }\textbf {\bibinfo
  {volume} {11}},\ \bibinfo {pages} {2621} (\bibinfo {year}
  {1997})}\BibitemShut {NoStop}%
\bibitem [{\citenamefont {Berry}(1984)}]{berry1984}%
  \BibitemOpen
  \bibfield  {author} {\bibinfo {author} {\bibfnamefont {M.~V.}\ \bibnamefont
  {Berry}},\ }\href {\doibase 10.1098/rspa.1984.0023} {\bibfield  {journal}
  {\bibinfo  {journal} {Proceedings of the Royal Society of London. A.
  Mathematical and Physical Sciences}\ }\textbf {\bibinfo {volume} {392}},\
  \bibinfo {pages} {45} (\bibinfo {year} {1984})}\BibitemShut {NoStop}%
\bibitem [{\citenamefont {Kramer}\ and\ \citenamefont
  {Saraceno}(1981)}]{kramer1981}%
  \BibitemOpen
  \bibfield  {author} {\bibinfo {author} {\bibfnamefont {P.}~\bibnamefont
  {Kramer}}\ and\ \bibinfo {author} {\bibfnamefont {M.}~\bibnamefont
  {Saraceno}},\ }\href@noop {} {{\selectlanguage {English}\emph {\bibinfo
  {title} {Geometry of the {{Time}}-{{Dependent Variational Principle}} in
  {{Quantum Mechanics}}}}}}\ (\bibinfo  {publisher} {{Springer}},\ \bibinfo
  {year} {1981})\BibitemShut {NoStop}%
\bibitem [{Note1()}]{Note1}%
  \BibitemOpen
  \bibinfo {note} {The correspondence is determined from our own numerical
  evaluation of the matrix element.}\BibitemShut {Stop}%
\bibitem [{\citenamefont {Wang}\ \emph {et~al.}(2017)\citenamefont {Wang},
  \citenamefont {Cooper}, \citenamefont {Halperin},\ and\ \citenamefont
  {Stern}}]{wang_particle-hole_2017}%
  \BibitemOpen
  \bibfield  {author} {\bibinfo {author} {\bibfnamefont {C.}~\bibnamefont
  {Wang}}, \bibinfo {author} {\bibfnamefont {N.~R.}\ \bibnamefont {Cooper}},
  \bibinfo {author} {\bibfnamefont {B.~I.}\ \bibnamefont {Halperin}}, \ and\
  \bibinfo {author} {\bibfnamefont {A.}~\bibnamefont {Stern}},\ }\href
  {\doibase 10.1103/PhysRevX.7.031029} {\bibfield  {journal} {\bibinfo
  {journal} {Phys. Rev. X}\ }\textbf {\bibinfo {volume} {7}},\ \bibinfo {pages}
  {031029} (\bibinfo {year} {2017})}\BibitemShut {NoStop}%
\bibitem [{\citenamefont {Pu}\ \emph {et~al.}(2018)\citenamefont {Pu},
  \citenamefont {Fremling},\ and\ \citenamefont {Jain}}]{pu_berry_2018}%
  \BibitemOpen
  \bibfield  {author} {\bibinfo {author} {\bibfnamefont {S.}~\bibnamefont
  {Pu}}, \bibinfo {author} {\bibfnamefont {M.}~\bibnamefont {Fremling}}, \ and\
  \bibinfo {author} {\bibfnamefont {J.~K.}\ \bibnamefont {Jain}},\ }\href
  {\doibase 10.1103/PhysRevB.98.075304} {\bibfield  {journal} {\bibinfo
  {journal} {Phys. Rev. B}\ }\textbf {\bibinfo {volume} {98}},\ \bibinfo
  {pages} {075304} (\bibinfo {year} {2018})}\BibitemShut {NoStop}%
\bibitem [{\citenamefont {Shao}\ \emph {et~al.}(2015)\citenamefont {Shao},
  \citenamefont {Kim}, \citenamefont {Haldane},\ and\ \citenamefont
  {Rezayi}}]{shao2015}%
  \BibitemOpen
  \bibfield  {author} {\bibinfo {author} {\bibfnamefont {J.}~\bibnamefont
  {Shao}}, \bibinfo {author} {\bibfnamefont {E.-A.}\ \bibnamefont {Kim}},
  \bibinfo {author} {\bibfnamefont {F.~D.~M.}\ \bibnamefont {Haldane}}, \ and\
  \bibinfo {author} {\bibfnamefont {E.~H.}\ \bibnamefont {Rezayi}},\ }\href
  {\doibase 10.1103/PhysRevLett.114.206402} {\bibfield  {journal} {\bibinfo
  {journal} {Phys. Rev. Lett.}\ }\textbf {\bibinfo {volume} {114}},\ \bibinfo
  {pages} {206402} (\bibinfo {year} {2015})}\BibitemShut {NoStop}%
\bibitem [{\citenamefont {Haldane}(2018)}]{haldane2018}%
  \BibitemOpen
  \bibfield  {author} {\bibinfo {author} {\bibfnamefont {F.~D.~M.}\
  \bibnamefont {Haldane}},\ }\href {\doibase 10.1063/1.5046122} {\bibfield
  {journal} {\bibinfo  {journal} {Journal of Mathematical Physics}\ }\textbf
  {\bibinfo {volume} {59}},\ \bibinfo {pages} {081901} (\bibinfo {year}
  {2018})}\BibitemShut {NoStop}%
\bibitem [{\citenamefont {Zak}(1964)}]{zak_magnetic_1964}%
  \BibitemOpen
  \bibfield  {author} {\bibinfo {author} {\bibfnamefont {J.}~\bibnamefont
  {Zak}},\ }\href {\doibase 10.1103/PhysRev.134.A1602} {\bibfield  {journal}
  {\bibinfo  {journal} {Phys. Rev.}\ }\textbf {\bibinfo {volume} {134}},\
  \bibinfo {pages} {A1602} (\bibinfo {year} {1964})}\BibitemShut {NoStop}%
\bibitem [{\citenamefont {Goldman}\ and\ \citenamefont
  {Fradkin}(2018)}]{goldman2018a}%
  \BibitemOpen
  \bibfield  {author} {\bibinfo {author} {\bibfnamefont {H.}~\bibnamefont
  {Goldman}}\ and\ \bibinfo {author} {\bibfnamefont {E.}~\bibnamefont
  {Fradkin}},\ }\href {\doibase 10.1103/PhysRevB.98.165137} {\bibfield
  {journal} {\bibinfo  {journal} {Phys. Rev. B}\ }\textbf {\bibinfo {volume}
  {98}},\ \bibinfo {pages} {165137} (\bibinfo {year} {2018})}\BibitemShut
  {NoStop}%
\bibitem [{\citenamefont {Thouless}(1984)}]{thouless1984}%
  \BibitemOpen
  \bibfield  {author} {\bibinfo {author} {\bibfnamefont {D.~J.}\ \bibnamefont
  {Thouless}},\ }\href {\doibase 10.1088/0022-3719/17/12/003} {\bibfield
  {journal} {\bibinfo  {journal} {Journal of Physics C: Solid State Physics}\
  }\textbf {\bibinfo {volume} {17}},\ \bibinfo {pages} {L325} (\bibinfo {year}
  {1984})}\BibitemShut {NoStop}%
\bibitem [{Note2()}]{Note2}%
  \BibitemOpen
  \bibinfo {note} {Note that for unnormalized wave functions, the Berry
  connection is determined by the formula $\protect \bm {A}_{\protect \bm
  {\alpha }}=-\protect \mathrm {Im}\mathinner {\langle {\Psi _{\protect \bm
  {\alpha }}|\partial _{\protect \bm {\alpha }}\Psi _{\protect \bm {\alpha
  }}}\rangle }/\mathinner {\langle {\Psi _{\protect \bm {\alpha }}|\Psi
  _{\protect \bm {\alpha }}}\rangle }$.}\BibitemShut {Stop}%
\bibitem [{\citenamefont {Pu}\ \emph {et~al.}(2017)\citenamefont {Pu},
  \citenamefont {Wu},\ and\ \citenamefont {Jain}}]{pu2017}%
  \BibitemOpen
  \bibfield  {author} {\bibinfo {author} {\bibfnamefont {S.}~\bibnamefont
  {Pu}}, \bibinfo {author} {\bibfnamefont {Y.-H.}\ \bibnamefont {Wu}}, \ and\
  \bibinfo {author} {\bibfnamefont {J.~K.}\ \bibnamefont {Jain}},\ }\href
  {\doibase 10.1103/PhysRevB.96.195302} {\bibfield  {journal} {\bibinfo
  {journal} {Phys. Rev. B}\ }\textbf {\bibinfo {volume} {96}},\ \bibinfo
  {pages} {195302} (\bibinfo {year} {2017})}\BibitemShut {NoStop}%
\bibitem [{\citenamefont {Haldane}(2004)}]{haldane2004}%
  \BibitemOpen
  \bibfield  {author} {\bibinfo {author} {\bibfnamefont {F.~D.~M.}\
  \bibnamefont {Haldane}},\ }\href {\doibase 10.1103/PhysRevLett.93.206602}
  {\bibfield  {journal} {\bibinfo  {journal} {Phys. Rev. Lett.}\ }\textbf
  {\bibinfo {volume} {93}},\ \bibinfo {pages} {206602} (\bibinfo {year}
  {2004})}\BibitemShut {NoStop}%
\bibitem [{\citenamefont {Jungwirth}\ \emph {et~al.}(2002)\citenamefont
  {Jungwirth}, \citenamefont {Niu},\ and\ \citenamefont
  {MacDonald}}]{jungwirth2002}%
  \BibitemOpen
  \bibfield  {author} {\bibinfo {author} {\bibfnamefont {T.}~\bibnamefont
  {Jungwirth}}, \bibinfo {author} {\bibfnamefont {Q.}~\bibnamefont {Niu}}, \
  and\ \bibinfo {author} {\bibfnamefont {A.~H.}\ \bibnamefont {MacDonald}},\
  }\href {\doibase 10.1103/PhysRevLett.88.207208} {\bibfield  {journal}
  {\bibinfo  {journal} {Phys. Rev. Lett.}\ }\textbf {\bibinfo {volume} {88}},\
  \bibinfo {pages} {207208} (\bibinfo {year} {2002})}\BibitemShut {NoStop}%
\bibitem [{\citenamefont {Ji}\ and\ \citenamefont
  {Shi}(2020)}]{ji_asymmetry_2020}%
  \BibitemOpen
  \bibfield  {author} {\bibinfo {author} {\bibfnamefont {G.}~\bibnamefont
  {Ji}}\ and\ \bibinfo {author} {\bibfnamefont {J.}~\bibnamefont {Shi}},\
  }\href {\doibase 10.1103/PhysRevB.101.235301} {\bibfield  {journal} {\bibinfo
   {journal} {Phys. Rev. B}\ }\textbf {\bibinfo {volume} {101}},\ \bibinfo
  {pages} {235301} (\bibinfo {year} {2020})}\BibitemShut {NoStop}%
\bibitem [{\citenamefont {Kamburov}\ \emph {et~al.}(2014)\citenamefont
  {Kamburov}, \citenamefont {Liu}, \citenamefont {Mueed}, \citenamefont
  {Shayegan}, \citenamefont {Pfeiffer}, \citenamefont {West},\ and\
  \citenamefont {Baldwin}}]{kamburov2014}%
  \BibitemOpen
  \bibfield  {author} {\bibinfo {author} {\bibfnamefont {D.}~\bibnamefont
  {Kamburov}}, \bibinfo {author} {\bibfnamefont {Y.}~\bibnamefont {Liu}},
  \bibinfo {author} {\bibfnamefont {M.~A.}\ \bibnamefont {Mueed}}, \bibinfo
  {author} {\bibfnamefont {M.}~\bibnamefont {Shayegan}}, \bibinfo {author}
  {\bibfnamefont {L.~N.}\ \bibnamefont {Pfeiffer}}, \bibinfo {author}
  {\bibfnamefont {K.~W.}\ \bibnamefont {West}}, \ and\ \bibinfo {author}
  {\bibfnamefont {K.~W.}\ \bibnamefont {Baldwin}},\ }\href {\doibase
  10.1103/PhysRevLett.113.196801} {\bibfield  {journal} {\bibinfo  {journal}
  {Phys. Rev. Lett.}\ }\textbf {\bibinfo {volume} {113}},\ \bibinfo {pages}
  {196801} (\bibinfo {year} {2014})}\BibitemShut {NoStop}%
\end{thebibliography}%

\end{document}